\documentclass[twocolumn,trackchanges,twocolappendix]{aastex631}
\usepackage{natbib}
\usepackage{epsfig}
\usepackage{graphicx}
\usepackage{subfigure}
\usepackage{float}
\usepackage{amsmath}
\usepackage{color}
\usepackage{amssymb}
\usepackage[utf8]{inputenc}
\usepackage{apjfonts}
\usepackage{hyperref}

\DeclareRobustCommand{\forceLower}[1]{\MakeLowercase{#1}}

\newcommand{\PMO}{Purple Mountain Observatory, Chinese Academy of Sciences, Nanjing 210023, China; jjwei@pmo.ac.cn, xfwu@pmo.ac.cn}
\newcommand{\USTC}{School of Astronomy and Space Sciences, University of Science and Technology of China, Hefei 230026, China}

\begin{document}

\title{Prospects for the Detection of High-Redshift Gamma-Ray Bursts in the Era of EP and SVOM}


\author[0000-0003-0162-2488]{Jun-Jie Wei}
\affiliation{\PMO}
\affiliation{\USTC}

\author[0000-0002-6299-1263]{Xue-Feng Wu}
\affiliation{\PMO}
\affiliation{\USTC}

\begin{abstract}
Gamma-ray bursts (GRBs) are a promising probe of the high-redshift Universe,
but their detection remains observationally challenging. In this work, we explore 
the detectability of high-$z$ GRBs by the Wide-field X-ray Telescope (WXT) aboard 
the Einstein Probe (\emph{EP}) and the coded-mask gamma-ray imager (ECLAIRs) aboard 
the Space-based multi-band astronomical Variable Objects Monitor (\emph{SVOM}).
Using a population synthesis model calibrated to $Swift$ GRB observations,
we develop a tool to estimate high-$z$ GRB detection rates for instruments 
with specific energy bands and sensitivities. Our results indicate that \emph{EP}/WXT
could detect $\sim5.1^{+3.4}_{-2.4}$ (with 68\% confidence level) GRBs annually 
at $z>6$, compared to $\sim0.7^{+1.0}_{-0.4}$ $\mathrm{events\,yr^{-1}}$ at $z>6$ 
for \emph{SVOM}/ECLAIRs. While \emph{EP} cannot independently determine redshifts 
(requiring optical/near-infrared follow-up), its assumed $\sim30\%$ follow-up
efficiency yields $\sim1.5^{+1.0}_{-0.7}$ confirmed $z>6$ GRBs annually.
\emph{SVOM}, equipped with dedicated follow-up telescopes, will promptly identify
high-$z$ candidates deserving deep near-infrared spectroscopy to ensure robust 
confirmation of high-$z$ GRBs. We anticipate that \emph{EP} and \emph{SVOM} will 
open new avenues for utilizing enlarged samples of high-$z$ GRBs to explore 
the early Universe. Moreover, \emph{EP} will assemble a substantial sample of soft,
low-luminosity GRBs at low-to-intermediate redshifts, providing critical
insights into the structure of GRB jets.
\end{abstract}


\section{Introduction}
\label{sec:intro}
Gamma-ray bursts (GRBs), among the most energetic explosive transients in our Universe,
are detectable up to very high redshifts. The most distant spectroscopically confirmed
burst to date is GRB 090423 at $z=8.2$ \citep{2009Natur.461.1258S,2009Natur.461.1254T}.
As such, GRBs represent a promising probe, complementary to quasar and galaxy surveys, 
for exploring the early Universe (see \citealt{2015JHEAp...7...35S,2015NewAR..67....1W}
for reviews). In particular, long GRBs with durations $T_{90}>2\,\mathrm{s}$
(where $T_{90}$ is defined as the time interval containing 90\% of the prompt emission; 
\citealt{1993ApJ...413L.101K}) arise from the core collapse of massive stars\footnote{Notably, 
two nearby long-duration bursts, GRB 211211A and GRB 230307A, exhibit kilonova signatures 
indicative of compact binary mergers, implying that at least some long GRBs originate from 
non-collapsar progenitors
\citep{2022Natur.612..223R,2022Natur.612..228T,2022Natur.612..232Y,2023ApJ...954L..29D,2024Natur.626..737L}.
Given that merging compact binary progenitors have significantly longer average lifetimes than 
collapsars, the fraction of merger-origin long-duration GRBs (especially those bright enough 
to be detectable) is expected to decrease substantially at high redshifts compared to $z=0$.} 
(e.g., \citealt{1993AAS...182.5505W,1998ApJ...494L..45P,2006ARA&A..44..507W}),
establishing them as powerful tracers of cosmic star formation 
(e.g., \citealt{1997ApJ...486L..71T,2007ApJ...671..272C,2008ApJ...683L...5Y,
2009ApJ...705L.104K,2009MNRAS.400L..10W,2012ApJ...744...95R,2013A&A...556A..90W,
2014MNRAS.439.3329W,2024ApJ...976L..16M}). 
Moreover, GRB afterglows are expected to be observable up to $z\sim20$ 
\citep{2000ApJ...540..687C,2004ApJ...604..508G}, 
enabling spectroscopic afterglow studies to provide independent constraints on 
the reionization history, ionizing photon escape fraction, and cosmic metal enrichment
history, and $\mathrm{21\,cm}$ absorption features \citep{2005ApJ...619..684I,
2006PASJ...58..485T,2014PASJ...66...63T,2012ApJ...760...27W,2013ApJ...774...26C,
2014arXiv1405.7400C,2015MNRAS.453..101C,2015A&A...580A.139H,2019MNRAS.483.5380T,
2023A&A...671A..84S,2025ApJ...985...28F,2025MNRAS.536.2839F,2025MNRAS.537..364S}.
GRBs may also constrain the stellar populations in the earliest galaxies by detecting
explosions from the first, metal-free stars (Population III stars;
\citealt{2006ApJ...642..382B,2010ApJ...715..967M,2011A&A...533A..32D,
2011ApJ...726..107S,2011ApJ...731..127T,2015ApJ...810...64M,2016ApJ...823...83M,
2019ApJ...878..128K}).

A major limitation is the extreme scarcity of high-redshift ($z\ge6$) GRB detections,
with only eleven events having been identified in 20 years of \emph{Swift} observations. 
Among these, seven have spectroscopic confirmations, while the remaining four rely on
photometric redshift estimates (as listed in Table~\ref{tab0}). This scarcity primarily 
results from current missions' limited sensitivity and relatively high-energy bandpass, 
as high-$z$ GRBs are faint with peak emission redshifted to lower energies. 
Although \emph{Swift} is suboptimal for detecting high-$z$ GRBs, previous studies 
suggest that $\sim2\%$ of its GRBs originate at $z>5.5$ \citep{2016ApJ...817....7P}. However, 
some high-$z$ events may have been missed during the patchy follow-up campaigns. Such patchiness 
arises when weather conditions prevent rapid follow-up observations, when afterglows are 
only observed at optical wavelengths rather than in the near-infrared (NIR), and because ground 
facilities with deep NIR spectroscopic capabilities for high-$z$ afterglows 
(requiring 8 m-class telescopes) remain scarce \citep{2022NatAs...6.1101C}.

\begin{table}
\renewcommand\arraystretch{1.1}
\tabcolsep=0.25cm
\centering \caption{List of high-redshift ($z\ge6$) GRBs detected by \emph{Swift}}
\begin{tabular}{lcccccc}
\hline
\hline
\multicolumn{3}{c}{Spectroscopic Redshifts}  &   & \multicolumn{3}{c}{Photometric Redshifts} \\
\cline{1-3} \cline{5-7}
  GRB & $z$ & Refs. &   & GRB & $z$ & Refs.\\
\hline
  050904 & 6.3 &  (1)  &  & 090429B & 9.4 & (12) \\
  080913 & 6.7 &  (2)  &  & 100905A & 7.9 & (13) \\
  090423 & 8.2 &  (3), (4) &  & 120521C & 6.0 & (14) \\
  130606A & 5.9 & (5), (6), (7)   &   &  120923A & 7.8 & (15)  \\
  140515A & 6.3 & (8), (9)   &   &      &     &  \\
  210905A & 6.3 & (10)   &    &     &     &  \\
  240218A & 6.8 & (11)   &    &     &     &  \\
\hline
\end{tabular}
\label{tab0}
\noindent{\footnotesize{\emph{References:} (1) \cite{2006Natur.440..184K}; (2) \cite{2009ApJ...693.1610G}; 
(3) \cite{2009Natur.461.1258S}; (4) \cite{2009Natur.461.1254T}; (5) \cite{2013ApJ...774...26C};
(6) \cite{2014PASJ...66...63T}; (7) \cite{2015A&A...580A.139H}; (8) \cite{2014arXiv1405.7400C};
(9) \cite{2015A&A...581A..86M}; (10) \cite{2023A&A...671A..84S}; (11) \cite{2024GCN.35756....1S};
(12) \cite{2011ApJ...736....7C}; (13) \cite{2018A&A...609A..62B}; (14) \cite{2014ApJ...781....1L}; 
(15) \cite{2018ApJ...865..107T}.}}
\end{table}

To fully leverage GRBs' potential for probing the early Universe, a much larger
statistical sample of high-$z$ GRBs is essential. Consequently, detecting high-$z$ 
GRBs drives the design of current and future space missions like the Einstein Probe 
(\emph{EP}; \citealt{2025arXiv250107362Y}) and the Space-based multi-band astronomical
Variable Objects Monitor (\emph{SVOM}; \citealt{wei2016}). GRB population studies
demonstrate that optimal detection of high-$z$ GRBs requires instruments operating 
in the soft X-ray band with high sensitivity 
\citep{Ghirlanda2015MNRAS,2015JHEAp...7...35S}. 

In this work, we explore the prospects for high-$z$ GRB detection during the \emph{EP}
and \emph{SVOM} observational era. \emph{EP} was launched in January 2024, followed by
\emph{SVOM} in June 2024; both missions have now commenced scientific operations.
Compared to other operational wide-field monitors, both the Wide-field X-ray Telescope
(WXT) on board \emph{EP} and the coded-mask gamma-ray imager (ECLAIRs) on board 
\emph{SVOM} feature high sensitivity with coverage extending into relatively 
low X-ray energies. For a 10-s exposure, WXT achieves a sensitivity of 
$\sim8.9\times10^{-10}\,\mathrm{erg\,cm^{-2}\,s^{-1}}$ 
in 0.5--4 keV, while ECLAIRs reaches 
$\sim7.2\times10^{-8}\,\mathrm{erg\,cm^{-2}\,s^{-1}}$ in 4--150 keV. These properties
make both instruments exceptionally well-suited for detecting high-$z$ GRBs.
Notably, \emph{EP}/WXT has already detected a GRB at $z=4.859$ (designated EP240315a; 
\citealt{2025NatAs...9..564L}), while \emph{SVOM}/ECLAIRs achieved the fifth-most
distant GRB redshift record with GRB 250314A at $z=7.3$ \citep{2025GCN.39732....1M}, 
demonstrating both missions' high-$z$ GRB survey capabilities.

This paper is structured as follows. Section~\ref{sec:method} details our methodology 
for computing high-$z$ GRB detection rates for instruments with specified energy bands 
and sensitivities. This approach builds on a population synthesis model calibrated to 
\emph{Swift} GRB observations. Section~\ref{sec:result} presents the projected 
populations of high-$z$ GRBs detectable by \emph{EP}/WXT and \emph{SVOM}/ECLAIRs. 
We conclude with a summary and discussion in Section~\ref{sec:conclusions}.
Throughout our analysis, we adopt a standard $\Lambda$CDM cosmology with $H_{0}=67.4$
$\mathrm{km\,s^{-1}\,Mpc^{-1}}$, $\Omega_{\rm m}=0.315$, and $\Omega_{\Lambda}=0.685$
\citep{2020AA...641A...6P}.

\section{Analysis Method}
\label{sec:method}
In this section, we describe the method to estimate the detection rate of high-$z$ GRBs for a generic detector
with specified energy band and sensitivity.

\subsection{The Population Model}
The differential rate of GRBs detected per unit time, within redshift $z$ to $z + dz$ and luminosity $L$ to 
$L + dL$, is given by
\begin{equation}
\Phi(L,\;z,\;t)=\frac{d^{3}N}{dtdzdL}= \frac{\Delta \Omega}{4\pi}\eta_{\rm duty}\frac{\psi(z)}{1+z}\frac{dV(z)}{dz}\phi(L)\;,
\end{equation}
where $\Delta \Omega$ is the solid angle of the sky covered by the detector, $\eta_{\rm duty}$ represents 
the detector's duty cycle (i.e., the fraction of time spent performing observations and searching for 
GRBs)\footnote{All satellites in low-Earth orbit are rendered inactive during passage through 
the South Atlantic Anomaly (SAA) due to elevated radiation levels. Additionally, spacecraft slewing 
accounts for a significant portion of the mission time \citep{2016ApJ...829....7L}.}, $\psi(z)$ denotes 
the comoving formation rate of GRBs (in units of $\mathrm{Gpc^{-3}\;yr^{-1}}$), the factor $(1+z)^{-1}$ 
accounts for cosmological time dilation, and $\phi(L)$ is the normalized luminosity function (LF) of GRBs. 
Finally, the comoving volume element in a flat $\Lambda$CDM cosmology is given by 
\begin{equation}
\frac{dV(z)}{dz}= \frac{4\pi cD_{L}^2(z)}{(1+z)^{2}H_{0}\sqrt{\Omega_{\mathrm{m}}(1+z)^{3}+\Omega_{\Lambda}}}\;,
\end{equation}
where $D_{L}(z)$ is the luminosity distance at $z$.

For an instrument with a flux threshold $P_{\rm lim}$, the number of GRBs detected with redshifts within 
$[0,\,z_{\rm max}]$ over an observing time $T$ can be expressed as
\begin{equation}
\label{eq:Nexp}
\begin{aligned}
N=\frac{\Delta \Omega}{4\pi}T\eta_{\rm duty}\int_{0}^{z_{\rm max}}\frac{\psi(z)}{1+z}\frac{dV(z)}{dz}dz
\int_{\max[L_{\rm min},\,L_{\rm lim}(z)]}^{L_{\rm max}}\phi(L)dL \,.
\end{aligned}
\end{equation}
Following \cite{Ghirlanda2015MNRAS}, we set the maximum redshift for our analysis to $z_{\rm max}=20$.
The LF is modeled to span a range of luminosities from $L_{\rm min}=10^{47}\,\mathrm{erg\,s^{-1}}$
to $L_{\rm max}=10^{55}\,\mathrm{erg\,s^{-1}}$. For a given redshift $z$, the luminosity threshold 
$L_{\rm lim}$, which appears in Equation~(\ref{eq:Nexp}), is calculated by equating the peak flux of 
the prompt GRB emission to the detector’s sensitivity threshold $P_{\rm lim}$, i.e.,
\begin{equation}
\label{eq:Llim}
L_{\rm lim}(z)=4\pi D_{L}^{2}(z)P_{\rm lim}k(z)\;,
\end{equation}
where $k(z)$ is the spectral $k$-correction factor that converts the observed flux in the detector's 
energy band $[E_{1},\,E_{2}]$ to the rest-frame flux of the burst in the $1$--$10^{4}$ keV band. 
For detectors with a photon flux threshold $P_{\rm lim}$ (in units of $\mathrm{ph\,cm^{-2}\,s^{-1}}$), 
the $k$-correction factor is defined as
\begin{equation}
\label{eq:k1}  
k(z)=\frac{\int_{1\,\mathrm{keV}/(1+z)}^{10^{4}\,\mathrm{keV}/(1+z)}EN(E)dE}{\int_{E_{1}}^{E_{2}}N(E)dE}\,,
\end{equation}
where $N(E)$ is the observed photon spectrum (more on this below). While for other detectors with 
an energy flux threshold $P_{\rm lim}$ (in units of $\mathrm{erg\,cm^{-2}\,s^{-1}}$), $k(z)$ should be 
rewritten as
\begin{equation}
\label{eq:k2} 
k(z)=\frac{\int_{1\,\mathrm{keV}/(1+z)}^{10^{4}\,\mathrm{keV}/(1+z)}EN(E)dE}{\int_{E_{1}}^{E_{2}}EN(E)dE}\,.
\end{equation}

\subsection{Simulation Setup}
\label{subsec:simulation}

\begin{table*}
\centering
\caption{The Properties of \emph{EP}/WXT and \emph{SVOM}/ECLAIRs}
\begin{tabular}{lccccc}
\hline
\hline
Mission  & Energy band  &  Sensitivity ($\texttt{@}10\,\mathrm{s}$)   &  Field of view   &  Duty cycle   & References\\
         &  ($\mathrm{keV}$)       &   ($\mathrm{erg\,cm^{-2}\,s^{-1}})$            &    ($\mathrm{sr}$)      &    & \\
\hline
\emph{EP}/$\mathrm{WXT}$        &  0.5--4      & $8.9 \times 10^{-10}$          &  1.1  &   67\%  & \cite{2025arXiv250107362Y}\\
\emph{SVOM}/$\mathrm{ECLAIRs}$  &  4--150      & $7.2\times 10^{-8}$            &  2.0  &  85\%  & \cite{wei2016}\\

\hline
\end{tabular}
\label{tab1}
\end{table*}

The expected detection rate of GRBs for a detector with defined energy range and sensitivity becomes 
calculable when the intrinsic GRB formation rate, $\psi(z)$, and LF, $\phi(L)$, in Equation~(\ref{eq:Nexp}) 
are specified. However, these parameters remain poorly 
constrained. A key objective of GRB population studies is to infer them through statistical fitting of the 
observed distributions of redshift, luminosity, and peak flux measurements
(e.g., \citealt{2001ApJ...548..522P,2002ApJ...574..554L,2019MNRAS.488.5823L,2004ApJ...611.1033F,2005ApJ...619..412G,
2005MNRAS.364L...8N,2006MNRAS.372.1034D,2007JCAP...07..003G,2007ApJ...661..394L,2007ApJ...656L..49S,
2008ApJ...673L.119K,2009ApJ...705L.104K,2009MNRAS.396..299S,Salvaterra2012ApJ,2010ApJ...711..495B,2010MNRAS.407.1972C,
2010MNRAS.406..558Q,2010MNRAS.406.1944W,2011MNRAS.416.2174C,2011MNRAS.417.3025V,2012ApJ...745..168L,
2012ApJ...744...95R,2013ApJ...772L...8T,2014MNRAS.439.3329W,2015ApJ...806...44P,2015MNRAS.454.1785T,
2015ApJS..218...13Y,2016ApJ...820...66D,2016A&A...587A..40P,2017IJMPD..2630002W,2018MNRAS.473.3385P,
2019MNRAS.488.4607L,2021MNRAS.508...52L,2020MNRAS.493.1479L,2021A&A...649A.166P,2022MNRAS.513.1078D,
2022ApJ...932...10G,2024ApJ...976..170Q,2025ApJ...982..148Q}). Such analyses typically adopt 
parameterized models for both the GRB formation rate and LF, and iteratively refine them by matching simulated 
populations to observational data. Although the specific functional forms vary across studies, a consensus exists 
that GRB populations likely underwent some kind of redshift evolution (e.g., 
\citealt{2009MNRAS.396..299S,Salvaterra2012ApJ,2011MNRAS.417.3025V,2019MNRAS.488.4607L,2021MNRAS.508...52L,
2022ApJ...932...10G}). Therefore, we employ several parameterized GRB rates and LFs, derived from 
recent representative population studies \citep{Salvaterra2012ApJ,2021MNRAS.508...52L,2022ApJ...932...10G}, 
to quantify the detection efficiency of high-$z$ GRBs for the \emph{EP}/WXT and \emph{SVOM}/ECLAIRs missions. 
Table~\ref{tab1} summarizes the energy bands, detection thresholds, fields of view, and duty cycles of
these instruments.

In this study, we first adopt the GRB formation rate and LF derived by \cite{Salvaterra2012ApJ},
hereafter \hyperlink{cite.Salvaterra2012ApJ}{S12}, as our baseline framework. To ensure robustness, 
we also perform analogous calculations using the parameterized functions from \cite{2021MNRAS.508...52L} 
and \cite{2022ApJ...932...10G}, finding similar results across all models (see Appendix~\ref{sec:appendix} 
for full comparative analysis).

Using the population synthesis framework from \hyperlink{cite.Salvaterra2012ApJ}{S12}, we calculate 
the expected detection rate of high-$z$ GRBs by \emph{EP} and \emph{SVOM}. The key assumptions adopted to 
simulate the burst population are summarized below (for full methodological details, see \citealt{Ghirlanda2015MNRAS}):

\begin{enumerate}
    \item  GRBs are distributed as a function of redshift (up to $z_{\rm max}=20$) according to 
    the GRB formation rate \citep{Salvaterra2012ApJ}:
    \begin{equation}
    \label{eq:GRBrate}
           \psi(z)\propto \psi_\star(z) (1+z)^{\delta}\;,
    \end{equation}
    where $\psi_\star(z)=(0.0157+0.118z)/[1+(z/3.23)^{4.66}]$ is the comoving cosmic star formation rate 
    (SFR), with units of ${\rm M}_{\odot}$ $\rm yr^{-1}$ $\rm Mpc^{-3}$ 
    \citep{Hopkins2006ApJ,Li2008MNRAS}.\footnote{Note that a proportionality coefficient relating $\psi(z)$ 
    (the GRB formation rate) to $\psi_\star(z)$ (the cosmic SFR) quantifies the GRB formation efficiency 
    per unit stellar mass, with units of ${\rm M}^{-1}_{\odot}$. This coefficient is determined by normalizing 
    the model predictions to the observed population of GRBs.} \hyperlink{cite.Salvaterra2012ApJ}{S12} 
    found that a strong redshift-dependent evolution 
    in the GRB rate density ($\delta=1.7\pm0.5$) is required to explain the observations. 

    \item The GRB LF is assumed as a broken power law \citep{Salvaterra2012ApJ}:
    \begin{equation}
          \phi(L) \propto \left\lbrace \begin{array}{ll}\left(\frac{L}{L_{b}}\right)^{-\nu_{1}}; ~~~~~~~L \leq L_{b} \\
                                          \left(\frac{L}{L_{b}}\right)^{-\nu_{2}}; ~~~~~~~L >L_{b}\;, \\
    \end{array} \right.
    \end{equation}
    where $\nu_{1}$ and $\nu_{2}$ are the power-law indices describing the LF below and above the break 
    luminosity $L_{b}$, respectively. \hyperlink{cite.Salvaterra2012ApJ}{S12} derived the best-fit parameters $\nu_{1}=1.50^{+0.16}_{-0.32}$, 
    $\nu_{2}=2.32^{+0.77}_{-0.32}$, and $L_{b}=(3.8^{+6.3}_{-2.7})\times10^{52}\,\mathrm{erg\,s^{-1}}$. 
    Here, the GRB LF is normalized to unity by integrating over the luminosity range 
    $L_{\rm min}=10^{47}\,\mathrm{erg\,s^{-1}}$ to $L_{\rm max}=10^{55}\,\mathrm{erg\,s^{-1}}$. 

    \item To compute the peak flux $P$ of GRBs in a given energy band, it is necessary to associate 
    a spectral model $N(E)$ to each mock burst characterized by redshift $z$ and luminosity $L$ 
    (see Equations~(\ref{eq:Llim})--(\ref{eq:k2})). Here we adopt a Band function spectrum \citep{Band1993ApJ}, 
    with the low-energy photon spectral index $\alpha$ and high-energy photon spectral index $\beta$ 
    sampled from Gaussian distributions centered at $-1$ and $-2.3$, respectively, both with 
    a standard deviation of $\sigma=0.2$ \citep{Kaneko2006ApJS,Nava2011A&A,2020ApJ...893...46V}.

    \item The observed peak energy $E_{\rm peak}$ in the Band spectrum is estimated from the empirical 
    $E_{\rm peak}$--$L$ correlation \citep{2004ApJ...609..935Y,Nava2012MNRAS}: 
    $\log_{10}[E_{\rm peak}(1+z)]=-25.33+0.53\log_{10}L$. The scatter of this correlation ($\sigma_{\rm sc}=0.29$ dex; 
    \citealt{Nava2012MNRAS}), which represents the standard deviation of the perpendicular distances of 
    the data points from the best-fit line, is incorporated into our simulations as the width of 
    the Gaussian distribution.
\end{enumerate}

\subsection{Simulation Data Calibration and Consistency Verification}
Our simulated GRB population is calibrated by the observed population of bright bursts detected by 
the Burst Alert Telescope (BAT) on board the \emph{Swift} satellite. 
\hyperlink{cite.Salvaterra2012ApJ}{S12} constructed a complete \emph{Swift}/BAT sample (the BAT6 sample), 
consisting of 58 GRBs selected under criteria favorable for redshift measurement \citep{2006A&A...447..897J} 
and with peak flux $P\ge2.6$ $\mathrm{\rm ph\,cm^{-2}\,s^{-1}}$ in the 15--150 keV energy range. 
Applying the same flux threshold used to define the BAT6 sample, we select 453 GRBs with 1-s peak flux
$P\ge2.6$ $\mathrm{\rm ph\,cm^{-2}\,s^{-1}}$ from the most recent \emph{Swift} sample (1425 GRBs with 
duration $T_{90}\ge2\,\mathrm{s}$, detected up to December
2024).\footnote{\url{https://swift.gsfc.nasa.gov/archive/grb_table/}}
Considering \emph{Swift}/BAT's field of view ($\mathrm{1.4\,sr}$), a mission period of $\sim20\,\mathrm{years}$, 
and an average instrument duty cycle of 78\% \citep{2016ApJ...829....7L}, the detection rate of GRBs 
with peak flux $P\ge2.6$ $\mathrm{\rm ph\,cm^{-2}\,s^{-1}}$ (15--150 keV) is estimated to be $\sim21$
$\mathrm{events\,yr^{-1}\,sr^{-1}}$. Our simulated GRB population is scaled to match this observed rate.

\begin{figure}
\begin{center}
\vskip-0.3in
\includegraphics[width=0.45\textwidth]{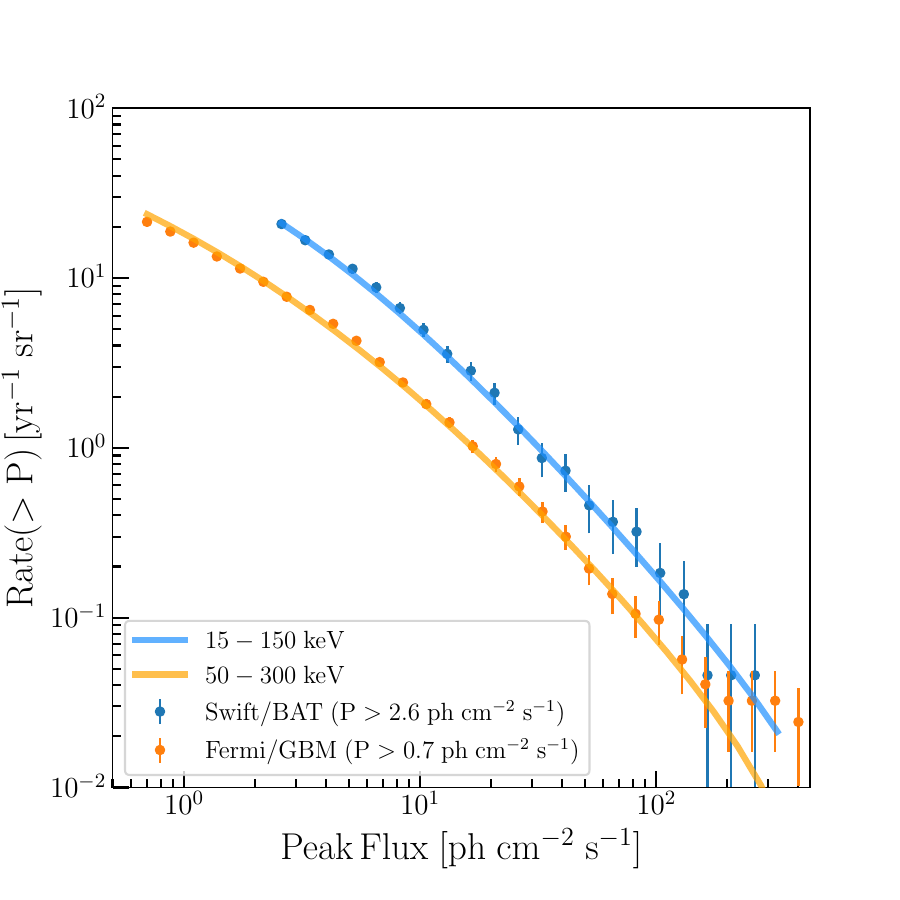}
\vskip-0.1in
\caption{Comparison of cumulative flux distributions between the simulated GRB population (solid lines)
and observed burst samples (dots). Blue and yellow dots represent the observed distributions for 
the \emph{Swift}/BAT (453 GRBs; $P\ge2.6$ $\mathrm{\rm ph\,cm^{-2}\,s^{-1}}$) and \emph{Fermi}/GBM 
(2637 GRBs; $P\ge0.7$ $\mathrm{\rm ph\,cm^{-2}\,s^{-1}}$) samples, respectively. \emph{Swift}/BAT 
fluxes are integrated in the 15--150 keV energy range, while \emph{Fermi}/GBM fluxes use 50--300 keV. 
Solid lines (color-matched to observations) show the simulated GRB population's flux distributions 
for each instrument's energy range. No attempt was made to fit the observed flux distributions.}
\label{fig1}
\vskip-0.2in
\end{center}
\end{figure}

Figure~\ref{fig1} displays the cumulative flux distribution of the observed \emph{Swift} sample (blue dots), 
comprising 453 GRBs with $P\ge2.6$ $\mathrm{\rm ph\,cm^{-2}\,s^{-1}}$. Using the simulation framework
described above, we compute the expected flux distribution for \emph{Swift}/BAT by adopting a detection
threshold of $P_{\rm lim}=2.6$ $\mathrm{\rm ph\,cm^{-2}\,s^{-1}}$ in the 15--150 keV energy band.
The simulated GRB population's flux distribution is shown as the blue solid line in Figure~\ref{fig1}.
We find that the expected flux distribution matches the observed \emph{Swift} sample well without requiring 
adjustment to the model parameters. This consistency supports the conclusion that the model assumptions 
(Section~\ref{subsec:simulation}), on which the code is based, are correctly implemented and successfully 
reproduce the population of \emph{Swift} bursts from which the model is derived.

To determine whether the simulated GRB population is also representative of samples observed by instruments 
beyond \emph{Swift}/BAT, we compare the cumulative flux distribution of the simulated population with that of 
the observed \emph{Fermi} sample. The public catalog of GRBs detected by \emph{Fermi}'s Gamma-ray Burst
Monitor (\emph{Fermi}/GBM) includes 3249 bursts with $T_{90}\ge2\,\mathrm{s}$, recorded between July 2008 
and December 2024 
\citep{2014ApJS..211...12G,2014ApJS..211...13V,2016ApJS..223...28N,2020ApJ...893...46V}.\footnote{\url{https://heasarc.gsfc.nasa.gov/W3Browse/fermi/fermigbrst.html}}
To address potential incompleteness in faint burst sampling, we consider the flux integrated over 
the 50--300 keV energy range and apply a cut to the GBM sample at $P\ge0.7$ $\mathrm{\rm ph\,cm^{-2}\,s^{-1}}$, 
resulting in 2637 GBM bursts. Considering an average sky coverage fraction of $\sim70\%$ for GBM,
a mission duration of $\sim16.5\,\mathrm{years}$ (encompassing our sample period), and an average instrument 
duty cycle of 85\% \citep{2020ApJ...893...46V}, the GBM detection rate for bursts with $P\ge0.7$ 
$\mathrm{\rm ph\,cm^{-2}\,s^{-1}}$ in the 50--300 keV energy range is $\sim21$ $\mathrm{events\,yr^{-1}\,sr^{-1}}$.
Figure~\ref{fig1} shows that the cumulative flux distribution of the 2637 observed \emph{Fermi} GRBs 
(yellow dots) aligns well with the corresponding distribution of the simulated GRB population (yellow 
solid line). Again, this agreement reinforces the reliability of our analysis.

\section{Populations of High-\forceLower{\emph{z}} GRBs Accessible by EP and SVOM}
\label{sec:result}
\begin{figure}
\begin{center}
\vskip-0.3in
\includegraphics[width=0.45\textwidth]{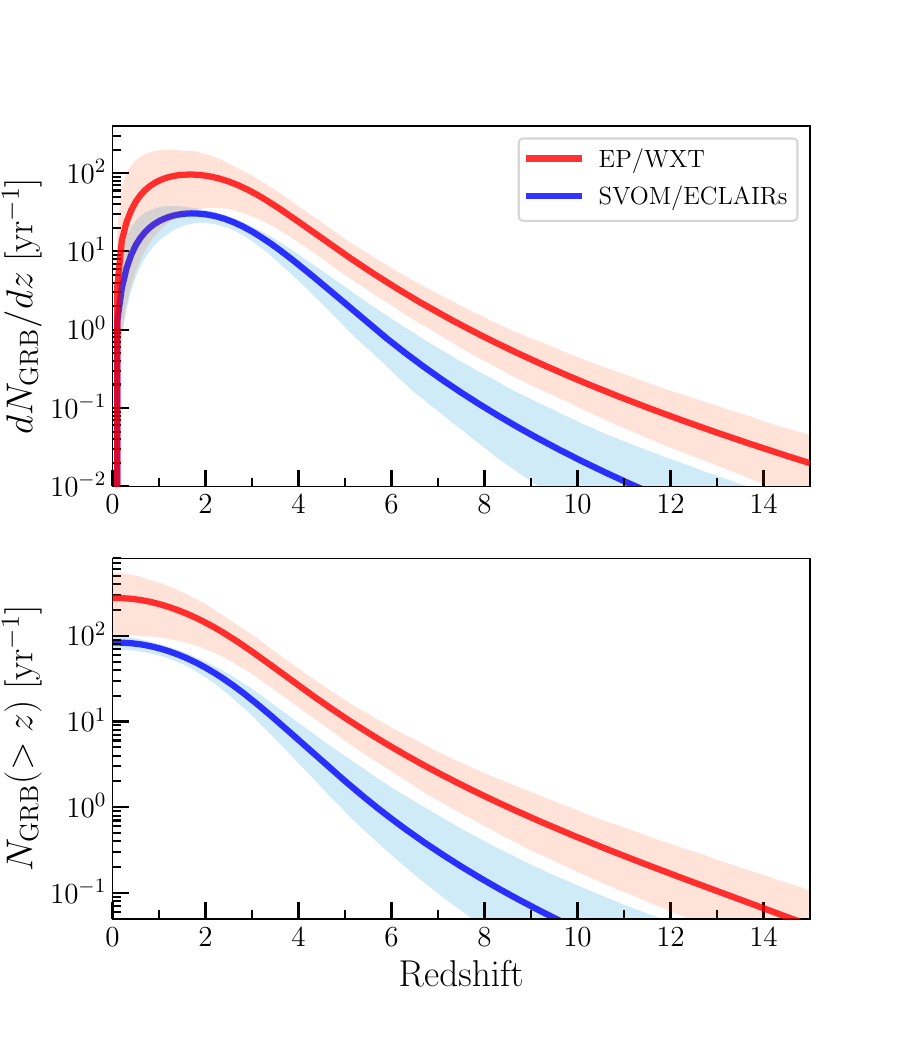}
\vskip-0.2in
\caption{Top panel: Redshift distributions of GRBs detectable by \emph{EP}/WXT (red curve) 
and \emph{SVOM}/ECLAIRs (blue curve), derived from the population synthesis framework of
\hyperlink{cite.Salvaterra2012ApJ}{S12}. Bottom panel: Cumulative detection rates for 
\emph{EP}/WXT and \emph{SVOM}/ECLAIRs, with color coding consistent with the top panel. 
Shaded regions, estimated by 1000 Monte Carlo simulations, represent $1\sigma$ confidence 
intervals incorporating uncertainties in the model parameters.} 
\label{fig2}
\vskip-0.2in
\end{center}
\end{figure}

Using the calibrated synthetic GRB population from \hyperlink{cite.Salvaterra2012ApJ}{S12}, we can predict 
the detection rate of an instrument operating in a specific energy band once its flux threshold is known. 
Figure~\ref{fig2} presents the redshift distributions and cumulative numbers of detected GRBs expected from 
observations by \emph{EP}/WXT and \emph{SVOM}/ECLAIRs (red curves for \emph{EP}/WXT and blue ones for 
\emph{SVOM}/ECLAIRs). The shaded regions around the model line represent the 68\% uncertainty in the model 
predictions. These uncertainties are estimated through 1000 Monte Carlo simulations incorporating uncertainties 
in the model parameters.

For \emph{EP}/WXT, we adopt a field of view of $\mathrm{1.1\,sr}$, a duty cycle of 67\%, and 
a flux sensitivity of $8.9\times 10^{-10}$ $\mathrm{\rm erg\,cm^{-2}\,s^{-1}}$ for a 10-s exposure 
in the 0.5--4 keV energy range \citep{2025arXiv250107362Y}. Within the models of \hyperlink{cite.Salvaterra2012ApJ}{S12},
the total number of GRB detections expected for a 1-year operation is $\simeq276^{+264}_{-175}$ 
($1\sigma$ confidence intervals; red curve in the bottom panel of Figure~\ref{fig2}), which may 
exceed the actual observed rate. As of May 2025, $\simeq90$ fast X-ray transients have been reported 
in General Coordinated Network (GCN) circulars,\footnote{\url{https://gcn.nasa.gov/circulars}} 
corresponding to $\simeq68$ $\mathrm{events\,yr^{-1}}$. However, not all detections are currently reported, 
and instrument characteristics remain under investigation. In addition, our choice of the limiting flux, 
$\sim8.9\times 10^{-10}$ $\mathrm{\rm erg\,cm^{-2}\,s^{-1}}$, may not be appropriate for comparing our 
estimates with the current observed detection count. This is because faint GRBs with peak fluxes slightly 
above the \emph{EP}/WXT threshold are not always reliably detected. Under the 0.5--4 keV sensitivity parameters 
of WXT, our implementation of the \hyperlink{cite.Salvaterra2012ApJ}{S12} framework yields high-$z$ GRB 
detection rates of $\sim5.1^{+3.4}_{-2.4}$ ($z>6$), $\sim1.3^{+1.2}_{-0.7}$ ($z>8$), and $\sim0.5^{+0.5}_{-0.3}$  
($z>10$) $\mathrm{events\,yr^{-1}}$, with uncertainties corresponding to $1\sigma$ confidence levels. 
These rates are listed in Columns 3-5 of Table~\ref{tab2}.

We implement \emph{SVOM}/ECLAIRs including a field of view of $\mathrm{2.0\,sr}$, 85\% duty cycle, and 4--150 keV
sensitivity threshold of $7.2\times 10^{-8}$ $\mathrm{\rm erg\,cm^{-2}\,s^{-1}}$ for a 10-s exposure \citep{wei2016}. 
Based on the models by \hyperlink{cite.Salvaterra2012ApJ}{S12}, the predicted total number for a 1-year operation 
is $\simeq83^{+16}_{-15}$ ($1\sigma$ confidence intervals; blue curve in the bottom panel of Figure~\ref{fig2}), 
roughly consistent with the observed rate. As of May 2025, $\simeq40$ GRBs have been detected by \emph{SVOM}/ECLAIRs 
(reported in GCN circulars), corresponding to $\simeq48$ $\mathrm{events\,yr^{-1}}$. We estimate that the detection 
rates of high-$z$ GRBs observable by \emph{SVOM}/ECLAIRs are $\sim0.7^{+1.0}_{-0.4}$, $\sim0.1^{+0.3}_{-0.1}$, 
and $\sim0.04^{+0.08}_{-0.03}$ $\mathrm{events\,yr^{-1}}$ at $z>6$, $z>8$, and $z>10$, respectively ($1\sigma$ 
confidence levels; see Columns 7-9 of Table~\ref{tab2}).

\begin{table*}
\renewcommand\arraystretch{1.3}
\tabcolsep=0.3cm
\centering \caption{Cumulative Detection Rates of GRBs by \emph{EP}/WXT 
and \emph{SVOM}/ECLAIRs Across Population-Synthesis Models}
\begin{tabular}{lccccccccc}
\hline
\hline
 & \multicolumn{9}{c}{Detection Rate $N_{\rm GRB}(>z)$ $(\mathrm{events\,yr^{-1}})$}   \\
 \cline{2-10} 
Model & \multicolumn{4}{c}{\emph{EP}/WXT}  &   & \multicolumn{4}{c}{\emph{SVOM}/ECLAIRs} \\
\cline{2-5} \cline{7-10}
  & $z>0$  & $z>6$ & $z>8$  & $z>10$ &   & $z>0$  & $z>6$ & $z>8$  & $z>10$\\
\hline
\cite{Salvaterra2012ApJ}  & $276^{+264}_{-175}$  & $5.1^{+3.4}_{-2.4}$  & $1.3^{+1.2}_{-0.7}$ & $0.5^{+0.5}_{-0.3}$ &   & $83^{+16}_{-15}$  & $0.7^{+1.0}_{-0.4}$ & $0.1^{+0.3}_{-0.1}$  & $0.04^{+0.08}_{-0.03}$\\
\cite{2021MNRAS.508...52L}  & $343^{+69}_{-54}$  & $3.5^{+0.9}_{-0.7}$  & $0.9^{+0.3}_{-0.2}$ & $0.3^{+0.1}_{-0.1}$ &   & $80^{+1}_{-1}$  & $0.7^{+0.2}_{-0.2}$ & $0.1^{+0.1}_{-0.1}$  & $0.02^{+0.02}_{-0.01}$\\
\cite{2022ApJ...932...10G} & $84^{+29}_{-24}$  & $4.7^{+4.3}_{-2.3}$  & $1.7^{+2.0}_{-1.0}$ & $0.7^{+1.0}_{-0.4}$ &   & $74^{+10}_{-11}$  & $1.3^{+2.1}_{-0.8}$ & $0.3^{+0.8}_{-0.2}$  & $0.11^{+0.31}_{-0.08}$\\

\hline
\end{tabular}
\label{tab2}
\end{table*}

\begin{figure}
\begin{center}
\includegraphics[width=0.45\textwidth]{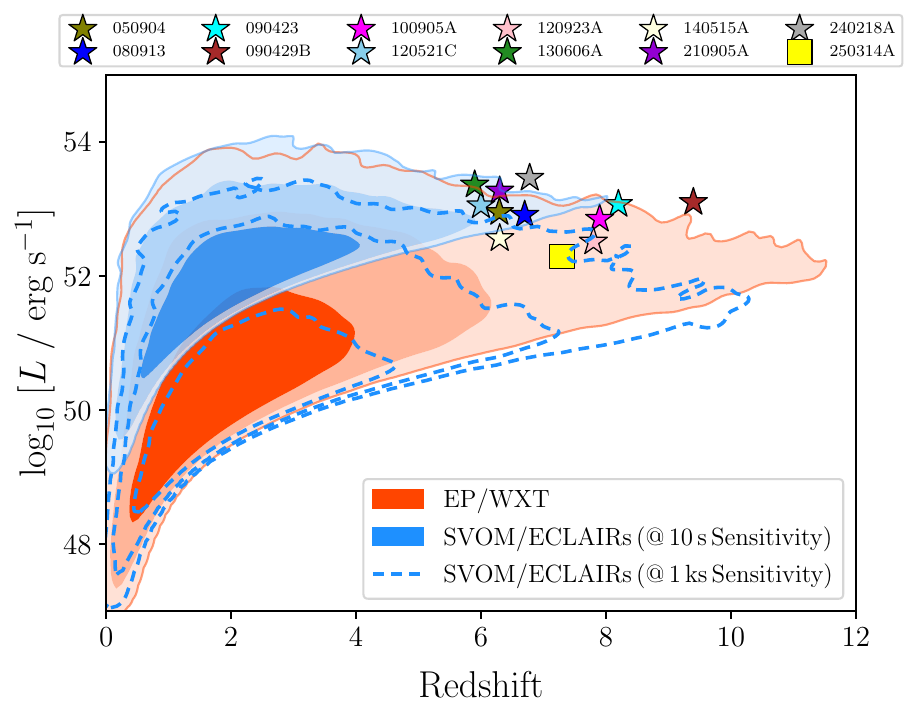}
\vskip-0.1in
\caption{Peak isotropic luminosity versus redshift of the simulated GRB populations detectable by \emph{EP}/WXT 
(red solid contours) and \emph{SVOM}/ECLAIRs (blue solid contours), with sensitivity thresholds for a 10-s
exposure. The shaded regions represent the $1-3\sigma$ confidence contours. The blue dashed contours correspond to 
\emph{SVOM}/ECLAIRs with a sensitivity threshold for a 1000-s exposure. The eleven high-redshift 
($z\geq6$) GRBs detected by \emph{Swift} are indicated by star symbols.
A square symbol marks the detection of long GRB 250314A at $z=7.3$ by \emph{SVOM}/ECLAIRs.} 
\label{fig3}
\vskip-0.2in
\end{center}
\end{figure}

By incorporating the properties of \emph{EP}/WXT and \emph{SVOM}/ECLAIRs (as listed in Table~\ref{tab1}), 
we run our population synthesis code to perform Monte Carlo simulations. Figure~\ref{fig3} displays 
the contour levels in the plane of the peak isotropic luminosity $L$ versus redshift $z$ for the simulated 
GRB populations detectable by \emph{EP}/WXT (red solid contours) and \emph{SVOM}/ECLAIRs (blue solid contours). 
For comparison, we include the eleven high-redshift ($z\geq6$) GRBs detected by \emph{Swift} 
(star symbols).\footnote{The relevant information for the peak luminosities of these eleven 
\emph{Swift} GRBs is available from published literature. For GRB 210905A, we used $L=1.87\times10^{53}$
$\mathrm{erg\,s^{-1}}$ from \cite{2022A&A...665A.125R}. For GRB 240218A, we calculated $L=2.88\times10^{53}$
$\mathrm{erg\,s^{-1}}$ using the peak photon flux and spectral parameters reported in \cite{2024GCN.35755....1V}. 
For the other nine GRBs, peak luminosities were taken from \cite{2021MNRAS.508...52L}.} 
Due to its extended energy range into soft X-rays and higher sensitivity, the \emph{EP}/WXT simulations sample 
relatively less energetic events compared to those currently detected by \emph{Swift} and those expected from 
\emph{SVOM} in the near future. GRB 250314A at $z=7.3$,\footnote{The peak luminosity of GRB 250314A is
$L=1.97\times10^{52}$ $\mathrm{erg\,s^{-1}}$, based on private communication with the \emph{SVOM} team.} detected by 
\emph{SVOM}/ECLAIRs (square symbol; \citealt{2025GCN.39732....1M}), lies outside the $3\sigma$ confidence region 
predicted by \emph{SVOM}/ECLAIRs simulations (blue solid contours). In our simulations,
the fiducial sensitivity of \emph{SVOM}/ECLAIRs is set to $7.2\times 10^{-8}$ $\mathrm{erg\,cm^{-2}\,s^{-1}}$, 
achievable with an exposure time of $\simeq10\,\mathrm{s}$. The instrument's maximal sensitivity, however, reaches
$7.2\times 10^{-10}$ $\mathrm{\rm erg\,cm^{-2}\,s^{-1}}$ for exposures of $\simeq1000\,\mathrm{s}$ \citep{wei2016}. 
Adopting this higher sensitivity threshold, the simulated GRB population detectable by \emph{SVOM}/ECLAIRs 
shifts to the blue dashed contours. Notably, GRB 250314A now falls within the $3\sigma$ confidence region of 
these updated predictions. In the 15--150 keV energy band, \emph{Swift}-detected long GRBs typically 
last tens to hundreds of seconds \citep{2016Ap&SS.361..155H}. At lower energies ($\mathrm{<10\,keV}$), they show 
substantially longer durations, with \emph{EP}-observed GRBs averaging several hundred seconds \citep{2025ApJ...986..106G}.
While cosmological time-dilation would prolong observed durations at high redshifts, bandpass-shifting and 
undetected weak prompt emission episodes counteract this effect \citep{2013MNRAS.436.3640L}. These factors 
collectively justify longer exposures for soft X/gamma-ray instruments like \emph{EP}/WXT and \emph{SVOM}/ECLAIRs.

\begin{figure}
\begin{center}
\includegraphics[width=0.45\textwidth]{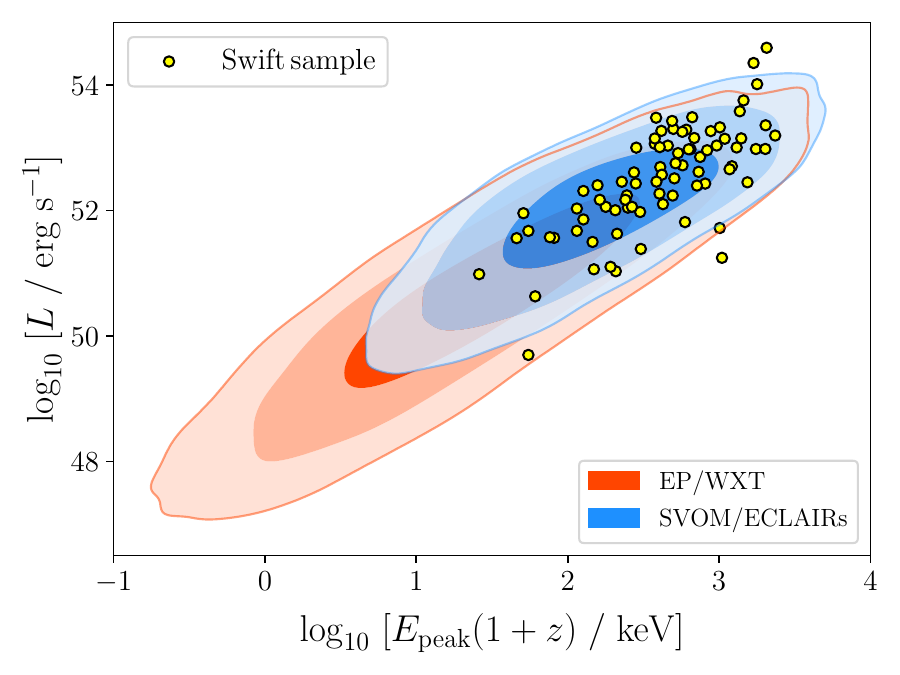}
\vskip-0.1in
\caption{Peak isotropic luminosity versus rest-frame peak energy of the simulated GRB populations detectable 
by \emph{EP}/WXT (red solid contours) and \emph{SVOM}/ECLAIRs (blue solid contours), with detection sensitivity 
thresholds corresponding to a 10-s exposure. The shaded regions represent the $1-3\sigma$ confidence contours. 
Yellow points indicate real GRBs detected by \emph{Swift} (adopted from \citealt{2016A&A...587A..40P}).} 
\label{fig4}
\vskip-0.2in
\end{center}
\end{figure}

Figure~\ref{fig4} compares the peak isotropic luminosity ($L$) versus rest-frame peak energy ($E_{\rm peak}(1+z)$)
of \emph{Swift}-detected GRBs (yellow dots) with simulated GRB populations detectable by \emph{EP}/WXT 
(red solid contours) and \emph{SVOM}/ECLAIRs (blue solid contours). The softer energy band and enhanced 
sensitivity of \emph{EP}/WXT enable it to detect a distinct GRB population dominated by soft spectra and 
low luminosities (primarily low-\emph{z} events), while simultaneously detecting high--\emph{z} GRBs 
whose spectra are softened by cosmological redshift effects. Compared to \emph{Swift} and \emph{SVOM}, 
\emph{EP}/WXT will advance the study of low-luminosity GRBs, which may exhibit different physical properties 
(e.g., opening angles or jet bulk velocities). Current knowledge of this population is limited to a few bursts 
due to the higher energy bands covered by existing missions. \emph{EP}/WXT's unique capability to trigger 
in the X-ray band will open a new window for studying low-$z$, soft-spectrum GRBs, which are predicted to
dominate the total GRB population. Based on our population models, we estimate that $\sim10\%$ of bursts 
detected at $z < 1$ will have luminosities less than $\mathrm{10^{49}\;erg\,s^{-1}}$ and soft spectra 
with peak energies $<20\,\mathrm{keV}$. A statistically robust sample of these soft, low-luminosity GRBs will 
provide critical insights into the nature of X-ray flashes, X-ray rich events \citep{2008ApJ...679..570S}, 
and low-luminosity GRBs \citep{2007ApJ...662.1111L,2015MNRAS.447.1911P,2016MNRAS.461.3607S}, whose origins 
remain unclear. Finally, the ratio of low-luminosity GRBs to supernovae could constrain the efficiency 
with which massive stars produce relativistic jets capable of breaking out of stellar envelopes \citep{2013MNRAS.428.1410G}.

\section{Summary and Discussions}\label{sec:conclusions}
In this work, we predict high-$z$ GRB detection rates for \emph{EP}/WXT and 
\emph{SVOM}/ECLAIRs. WXT's soft X-ray band (0.5--4 keV) and exceptional sensitivity
($\sim8.9\times10^{-10}\,\mathrm{erg\,cm^{-2}\,s^{-1}}$) make it ideally suited for
identifying high-$z$ GRBs, as demonstrated by its detection of EP240315a at $z=4.859$
\citep{2025NatAs...9..564L}. ECLAIRs' 4 keV low energy threshold similarly facilitates 
high-$z$ detections, exemplified by GRB 250314A at $z=7.3$ \citep{2025GCN.39732....1M}. 
Based on a physically motivated population synthesis model calibrated to \emph{Swift} 
observations, we evaluate the redshift distributions and detection rates of GRBs 
for both instruments. Our results indicate that \emph{EP}/WXT could detect
$\sim5.1^{+3.4}_{-2.4}$ GRBs at $z>6$ for a 1 yr operation, while \emph{SVOM}/ECLAIRs 
is expected to detect $\sim0.7^{+1.0}_{-0.4}$ $\mathrm{events\,yr^{-1}}$ at $z>6$.
In addition to the representative GRB population model of 
\hyperlink{cite.Salvaterra2012ApJ}{S12}, we perform analogous calculations using 
alternative population synthesis models from \cite{2021MNRAS.508...52L} and
\cite{2022ApJ...932...10G}. We derive consistent high-$z$ detection rates across 
all models, demonstrating robustness to population synthesis assumptions.

Note that \emph{EP}, operating solely as an X-ray telescope, cannot determine redshifts.
Optical/NIR follow-up observations are essential for identifying 
high-$z$ GRBs, though such follow-ups may not always be feasible for \emph{EP}-detected 
events. This could yield fewer confirmed high-$z$ GRBs than our estimates suggest.
Assuming a $\sim30\%$ efficiency for ground-based spectroscopic redshift determination, 
we expect \emph{EP} to detect $\sim1.5^{+1.0}_{-0.7}$ GRBs at $z>6$ annually. 
Equipped with dedicated follow-up telescopes, \emph{SVOM} will play a crucial role 
in promptly identifying high-$z$ GRB candidates deserving deep NIR spectroscopy 
(see \citealt{2024A&A...685A.163L}). Specifically, very high-redshift cases may be 
indicated by non-detection in both channels of the on-board Visible Telescope (VT)  
(though not uniquely, as dusty GRBs would also appear optically faint). These candidates 
can then be prioritized for ground-based NIR follow-up observations using positional locations 
from \emph{SVOM}'s Microchannel X-ray Telescope (MXT).

After several years of operation, we anticipate that \emph{EP} and \emph{SVOM} will 
double the number of existing high-$z$ GRBs. Furthermore, their substantial sample 
of low-luminosity GRBs at low-to-intermediate redshifts will enable investigations 
into potential sub-populations with dominant soft X-ray flashes and the faint end 
of the GRB LF---both offering key insights into GRB jet structure 
\citep{2007ApJ...662.1111L,2015MNRAS.447.1911P,2016MNRAS.461.3607S}.
In conclusions, \emph{EP} and \emph{SVOM} will open new avenues for advancing GRB
studies and early Universe exploration through high-$z$ beacons that probe the cosmic 
frontier. Notably, besides \emph{EP} and \emph{SVOM}, several mission concepts 
specifically designed for high-$z$ GRB discovery, including THESEUS 
\citep{2018AdSpR..62..191A,2021ExA....52..183A,2021ExA....52..277G}, Gamow Explorer
\citep{2020grbg.conf...51W}, and HiZ-GUNDAM \citep{2024SPIE13093E..20Y}, are under 
active investigation.

\section*{Acknowledgments}
We would like to thank the anonymous referee for valuable comments that helped improve 
our manuscript. We are also grateful to Giancarlo Ghirlanda, Hou-Dun Zeng, Guang-Xuan Lan, 
Chong-Yu Gao, Chen-Wei Wang, and Wen-Long Zhang for useful discussions.
This work is supported by the Strategic Priority Research Program of the Chinese Academy
of Sciences (grant No. XDB0550400), the National Key R\&D Program of China (2024YFA1611704),
the National Natural Science Foundation of China (grant Nos. 12422307, 12373053, and 12321003),
the Key Research Program of Frontier Sciences (grant No. ZDBS-LY-7014) of Chinese Academy of
Sciences, and the Natural Science Foundation of Jiangsu Province (grant No. BK20221562).

\appendix

\section{Alternative GRB Formation Rate and LF}
\label{sec:appendix}
We briefly discuss results for alternative GRB formation rates and LFs. Alongside the representative
GRB population model of \hyperlink{cite.Salvaterra2012ApJ}{S12}, we consider analyses from 
\cite{2021MNRAS.508...52L} and \cite{2022ApJ...932...10G}.

\begin{figure*}
\begin{center}
\vskip-0.2in
\includegraphics[width=0.45\textwidth]{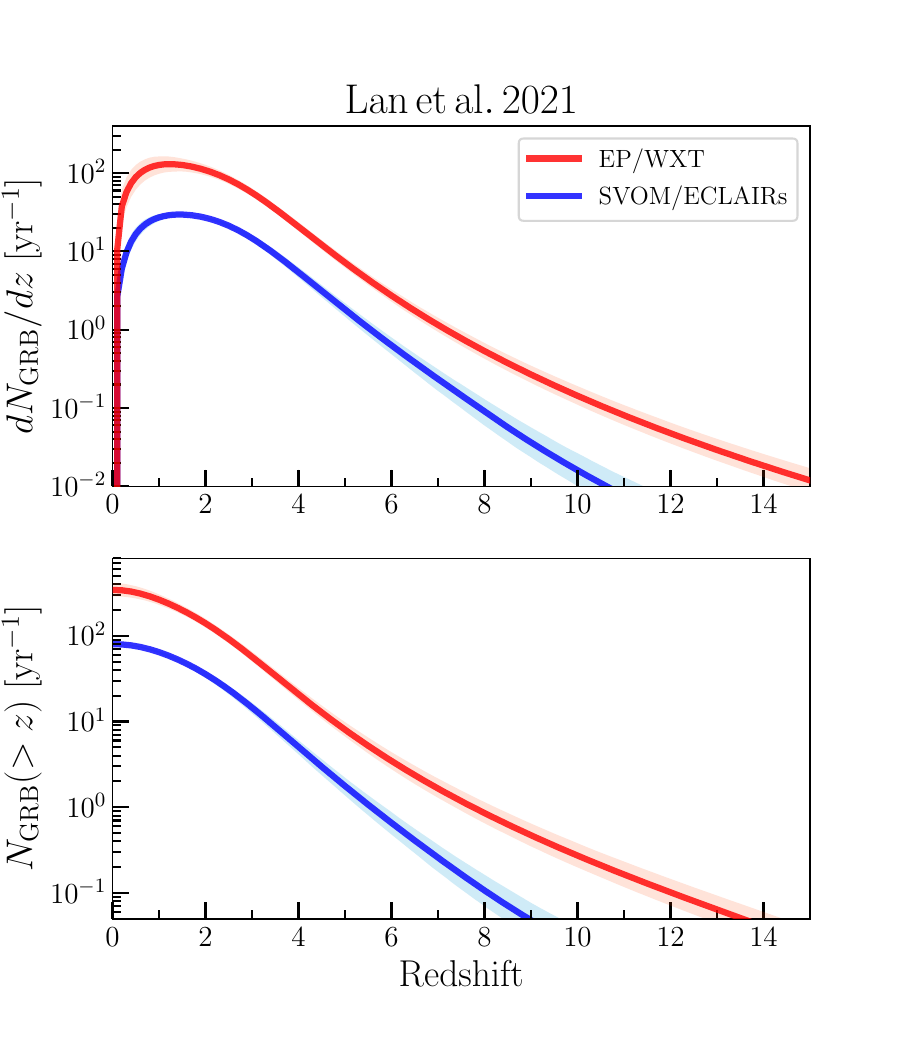}
\includegraphics[width=0.45\textwidth]{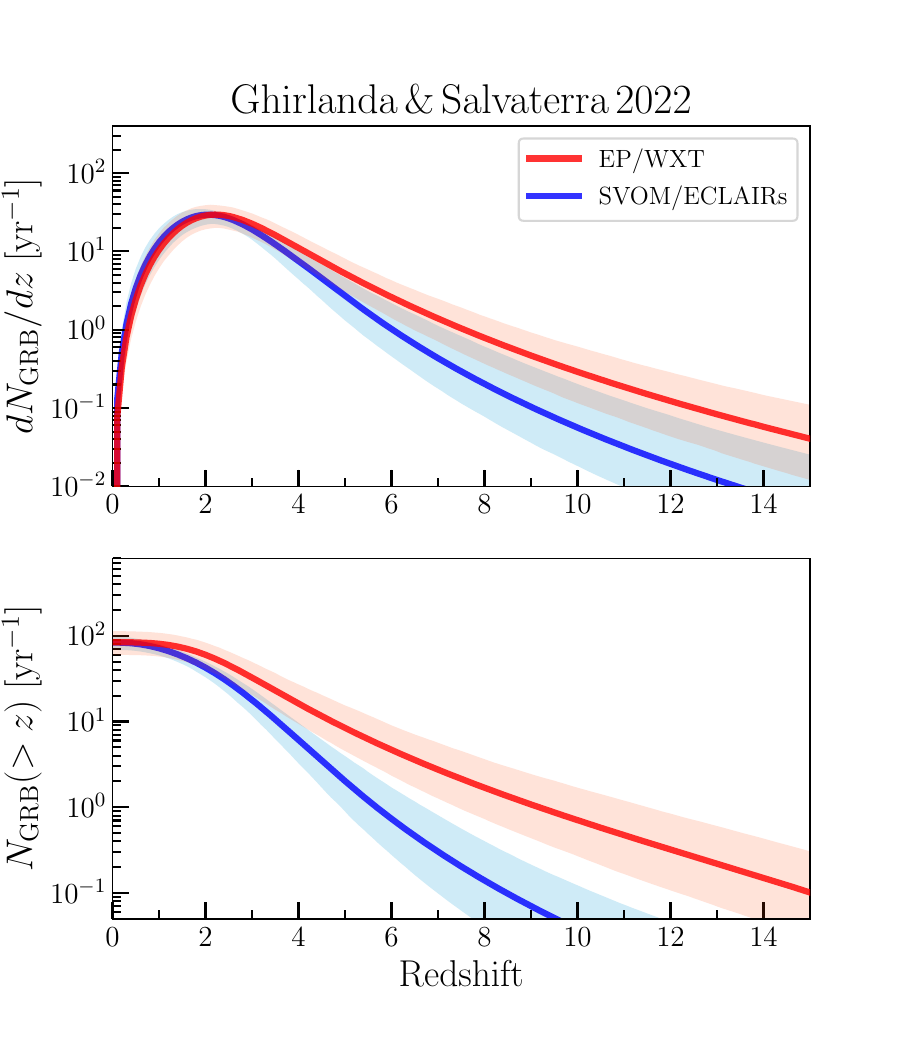}
\vskip-0.2in
\caption{Same as Figure~\ref{fig2}, but now for the GRB formation rate and LF models of \cite{2021MNRAS.508...52L}
and \cite{2022ApJ...932...10G}.} 
\label{fig5}
\vskip-0.2in
\end{center}
\end{figure*}

\cite{2021MNRAS.508...52L} analyzed 302 GRBs detected by \emph{Swift} up to November 2019 
with peak flux $P\ge1.0$ $\mathrm{\rm ph\,cm^{-2}\,s^{-1}}$ in the 15--150 keV energy range. 
They considered three different scenarios: (i) both a non-evolving GRB formation rate
(where the GRB formation rate strictly follows the cosmic SFR) and a non-evolving LF; 
(ii) a redshift-evolving GRB formation rate with a non-evolving LF; and (iii) a non-evolving 
GRB formation rate with a redshift-evolving LF. Here we adopt their results for scenario (ii). 
That is, the GRB formation rate is proportional to the SFR with an evolutionary term parameterized 
by $(1+z)^\delta$, as in Equation~(\ref{eq:GRBrate}), and the non-evolving LF is described by 
a broken power law:
\begin{equation}
          \phi(L) = \frac{A}{\ln(10)L}\left\lbrace \begin{array}{ll}\left(\frac{L}{L_{b}}\right)^{-\nu_{1}}; ~~~~~~~L \leq L_{b} \\
                                          \left(\frac{L}{L_{b}}\right)^{-\nu_{2}}; ~~~~~~~L >L_{b}\;, \\
    \end{array} \right.
    \end{equation}
where $A$ is a normalization constant.\footnote{To maintain consistency with the 
\hyperlink{cite.Salvaterra2012ApJ}{S12} model, $A$ is determined by normalizing the LF such that 
its integral equals unity over the luminosity range $L_{\rm min}=10^{47}\,\mathrm{erg\,s^{-1}}$ 
to $L_{\rm max}=10^{55}\,\mathrm{erg\,s^{-1}}$.} The model parameters are obtained as 
$\delta=1.43^{+0.22}_{-0.20}$, $\nu_{1}=0.60^{+0.05}_{-0.05}$, $\nu_{2}=1.65^{+0.27}_{-0.28}$, and 
$\log_{10}\left(L_{b}/\,\mathrm{erg\,s^{-1}}\right)=52.98^{+0.11}_{-0.12}$. We extracted 942 GRBs 
with peak flux $P\ge1.0$ $\mathrm{\rm ph\,cm^{-2}\,s^{-1}}$ (i.e., the same flux threshold used to 
define the sample of \cite{2021MNRAS.508...52L}) from the most recent \emph{Swift} sample of 1425 GRBs 
(all with $T_{90}\ge2\,\mathrm{s}$) detected through December 2024. Considering a field of view of 
$\mathrm{1.4\,sr}$, a mission period of $\sim20\,\mathrm{years}$, and an average instrument duty cycle 
of 78\%, the \emph{Swift}/BAT detection rate for bursts with peak flux $P\ge1.0$ $\mathrm{\rm ph\,cm^{-2}\,s^{-1}}$ 
in the 15--150 keV band is $\sim43$ $\mathrm{events\,yr^{-1}\,sr^{-1}}$. Within the population 
synthesis framework of \cite{2021MNRAS.508...52L}, we calibrate our simulated GRB population to 
this observed rate.

\cite{2022ApJ...932...10G} analyzed the BAT6 sample, which consists of GRBs with a 15--150 keV peak flux
$P\ge2.6$ $\mathrm{\rm ph\,cm^{-2}\,s^{-1}}$. They parameterized the GRB formation rate as:
\begin{equation}
     \psi(z)\propto \frac{(1+z)^{q_{1}}}{1+\left(\frac{1+z}{q_{2}}\right)^{q_{3}}}\;,
\end{equation}
a functional form motivated by its established use for fitting the cosmic SFR \citep{2014ARA&A..52..415M}.
Their LF adopts a broken power-law form defined for $L\ge10^{47}\,\mathrm{erg\,s^{-1}}$ with 
a redshift-evolving break luminosity: 
\begin{equation}
          \phi(L,\,z) \propto\left\lbrace \begin{array}{ll}\left(\frac{L}{L_{b,z}}\right)^{-\nu_{1}}; ~~~~~~~L \leq L_{b,z} \\
                                          \left(\frac{L}{L_{b,z}}\right)^{-\nu_{2}}; ~~~~~~~L >L_{b,z}\;, \\
    \end{array} \right.
    \end{equation}
where $L_{b,z}=L_{b,0}(1+z)^\delta$. The model parameters are obtained as $q_{1}=3.33^{+0.33}_{-0.33}$,
$q_{2}=3.42^{+0.28}_{-0.28}$, $q_{3}=6.21^{+0.38}_{-0.32}$, $\nu_{1}=0.97^{+0.05}_{-0.04}$, 
$\nu_{2}=2.21^{+0.13}_{-0.18}$, $\log_{10}\left(L_{b,0}/\,\mathrm{erg\,s^{-1}}\right)=52.02^{+0.22}_{-0.19}$, 
and $\delta=0.64^{+0.32}_{-0.26}$. Within the population synthesis framework of \cite{2022ApJ...932...10G}, 
we adopt the same \emph{Swift}/BAT detection rate ($\sim21$ $\mathrm{events\,yr^{-1}\,sr^{-1}}$) for bursts 
with peak flux $P\ge2.6$ $\mathrm{\rm ph\,cm^{-2}\,s^{-1}}$ in the 15--150 keV band as used in the 
\hyperlink{cite.Salvaterra2012ApJ}{S12} model to calibrate our simulated GRB population.

Figure~\ref{fig5} shows the redshift distributions and cumulative numbers of GRBs detectable by 
\emph{EP}/WXT and \emph{SVOM}/ECLAIRs under the \cite{2021MNRAS.508...52L} and \cite{2022ApJ...932...10G} 
models. The expected GRB detection rates at various redshifts by \emph{EP}/WXT and \emph{SVOM}/ECLAIRs 
for these models are summarized in Table~\ref{tab2}. Although the adopted GRB formation rates and LFs 
differ from those in the \hyperlink{cite.Salvaterra2012ApJ}{S12} model, the high-$z$ GRB detection rates
show qualitative similarities.


\begin{thebibliography}{}
\expandafter\ifx\csname natexlab\endcsname\relax\def\natexlab#1{#1}\fi
\providecommand{\url}[1]{\href{#1}{#1}}
\providecommand{\dodoi}[1]{doi:~\href{http://doi.org/#1}{\nolinkurl{#1}}}
\providecommand{\doeprint}[1]{\href{http://ascl.net/#1}{\nolinkurl{http://ascl.net/#1}}}
\providecommand{\doarXiv}[1]{\href{https://arxiv.org/abs/#1}{\nolinkurl{https://arxiv.org/abs/#1}}}

\bibitem[{{Amati} {et~al.}(2018){Amati}, {O'Brien}, {G{\"o}tz}, {Bozzo},
  {Tenzer}, {Frontera}, {Ghirlanda}, {Labanti}, {Osborne}, {Stratta}, {Tanvir},
  {Willingale}, {Attina}, {Campana}, {Castro-Tirado}, {Contini}, {Fuschino},
  {Gomboc}, {Hudec}, {Orleanski}, {Renotte}, {Rodic}, {Bagoly}, {Blain},
  {Callanan}, {Covino}, {Ferrara}, {Le Floch}, {Marisaldi}, {Mereghetti},
  {Rosati}, {Vacchi}, {D'Avanzo}, {Giommi}, {Piranomonte}, {Piro}, {Reglero},
  {Rossi}, {Santangelo}, {Salvaterra}, {Tagliaferri}, {Vergani}, {Vinciguerra},
  {Briggs}, {Campolongo}, {Ciolfi}, {Connaughton}, {Cordier}, {Morelli},
  {Orlandini}, {Adami}, {Argan}, {Atteia}, {Auricchio}, {Balazs}, {Baldazzi},
  {Basa}, {Basak}, {Bellutti}, {Bernardini}, {Bertuccio}, {Braga}, {Branchesi},
  {Brandt}, {Brocato}, {Budtz-Jorgensen}, {Bulgarelli}, {Burderi}, {Camp},
  {Capozziello}, {Caruana}, {Casella}, {Cenko}, {Chardonnet}, {Ciardi},
  {Colafrancesco}, {Dainotti}, {D'Elia}, {De Martino}, {De Pasquale}, {Del
  Monte}, {Della Valle}, {Drago}, {Evangelista}, {Feroci}, {Finelli},
  {Fiorini}, {Fynbo}, {Gal-Yam}, {Gendre}, {Ghisellini}, {Grado}, {Guidorzi},
  {Hafizi}, {Hanlon}, {Hjorth}, {Izzo}, {Kiss}, {Kumar}, {Kuvvetli}, {Lavagna},
  {Li}, {Longo}, {Lyutikov}, {Maio}, {Maiorano}, {Malcovati}, {Malesani},
  {Margutti}, {Martin-Carrillo}, {Masetti}, {McBreen}, {Mignani}, {Morgante},
  {Mundell}, {Nargaard-Nielsen}, {Nicastro}, {Palazzi}, {Paltani}, {Panessa},
  {Pareschi}, {Pe'er}, {Penacchioni}, {Pian}, {Piedipalumbo}, {Piran}, {Rauw},
  {Razzano}, {Read}, {Rezzolla}, {Romano}, {Ruffini}, {Savaglio}, {Sguera},
  {Schady}, {Skidmore}, {Song}, {Stanway}, {Starling}, {Topinka}, {Troja}, {van
  Putten}, {Vanzella}, {Vercellone}, {Wilson-Hodge}, {Yonetoku}, {Zampa},
  {Zampa}, {Zhang}, {Zhang}, {Zhang}, {Zhang}, {Antonelli}, {Bianco}, {Boci},
  {Boer}, {Botticella}, {Boulade}, {Butler}, {Campana}, {Capitanio}, {Celotti},
  {Chen}, {Colpi}, {Comastri}, {Cuby}, {Dadina}, {De Luca}, {Dong}, {Ettori},
  {Gandhi}, {Geza}, {Greiner}, {Guiriec}, {Harms}, {Hernanz}, {Hornstrup},
  {Hutchinson}, {Israel}, {Jonker}, {Kaneko}, {Kawai}, {Wiersema}, {Korpela},
  {Lebrun}, {Lu}, {MacFadyen}, {Malaguti}, {Maraschi}, {Melandri}, {Modjaz},
  {Morris}, {Omodei}, {Paizis}, {P{\'a}ta}, {Petrosian}, {Rachevski}, {Rhoads},
  {Ryde}, \& {Sabau-Graziati}}]{2018AdSpR..62..191A}
{Amati}, L., {O'Brien}, P., {G{\"o}tz}, D., {et~al.} 2018, Advances in Space
  Research, 62, 191, \dodoi{10.1016/j.asr.2018.03.010}

\bibitem[{{Amati} {et~al.}(2021){Amati}, {O'Brien}, {G{\"o}tz}, {Bozzo},
  {Santangelo}, {Tanvir}, {Frontera}, {Mereghetti}, {Osborne}, {Blain}, {Basa},
  {Branchesi}, {Burderi}, {Caballero-Garc{\'\i}a}, {Castro-Tirado},
  {Christensen}, {Ciolfi}, {De Rosa}, {Doroshenko}, {Ferrara}, {Ghirlanda},
  {Hanlon}, {Heddermann}, {Hutchinson}, {Labanti}, {Le Floch}, {Lerman},
  {Paltani}, {Reglero}, {Rezzolla}, {Rosati}, {Salvaterra}, {Stratta},
  {Tenzer}, \& {Theseus Consortium}}]{2021ExA....52..183A}
{Amati}, L., {O'Brien}, P.~T., {G{\"o}tz}, D., {et~al.} 2021, Experimental
  Astronomy, 52, 183, \dodoi{10.1007/s10686-021-09807-8}

\bibitem[{{Band} {et~al.}(1993){Band}, {Matteson}, {Ford}, {Schaefer},
  {Palmer}, {Teegarden}, {Cline}, {Briggs}, {Paciesas}, {Pendleton}, {Fishman},
  {Kouveliotou}, {Meegan}, {Wilson}, \& {Lestrade}}]{Band1993ApJ}
{Band}, D., {Matteson}, J., {Ford}, L., {et~al.} 1993, \apj, 413, 281,
  \dodoi{10.1086/172995}

\bibitem[{{Bolmer} {et~al.}(2018){Bolmer}, {Greiner}, {Kr{\"u}hler}, {Schady},
  {Ledoux}, {Tanvir}, \& {Levan}}]{2018A&A...609A..62B}
{Bolmer}, J., {Greiner}, J., {Kr{\"u}hler}, T., {et~al.} 2018, \aap, 609, A62,
  \dodoi{10.1051/0004-6361/201731255}

\bibitem[{{Bromm} \& {Loeb}(2006)}]{2006ApJ...642..382B}
{Bromm}, V., \& {Loeb}, A. 2006, \apj, 642, 382, \dodoi{10.1086/500799}

\bibitem[{{Butler} {et~al.}(2010){Butler}, {Bloom}, \&
  {Poznanski}}]{2010ApJ...711..495B}
{Butler}, N.~R., {Bloom}, J.~S., \& {Poznanski}, D. 2010, \apj, 711, 495,
  \dodoi{10.1088/0004-637X/711/1/495}

\bibitem[{{Campana} {et~al.}(2022){Campana}, {Ghirlanda}, {Salvaterra},
  {Gonzalez}, {Landoni}, {Pariani}, {Riva}, {Riva}, {Smartt}, {Tanvir}, \&
  {Vergani}}]{2022NatAs...6.1101C}
{Campana}, S., {Ghirlanda}, G., {Salvaterra}, R., {et~al.} 2022, Nature
  Astronomy, 6, 1101, \dodoi{10.1038/s41550-022-01804-x}

\bibitem[{{Campisi} {et~al.}(2010){Campisi}, {Li}, \&
  {Jakobsson}}]{2010MNRAS.407.1972C}
{Campisi}, M.~A., {Li}, L.~X., \& {Jakobsson}, P. 2010, \mnras, 407, 1972,
  \dodoi{10.1111/j.1365-2966.2010.17044.x}

\bibitem[{{Cao} {et~al.}(2011){Cao}, {Yu}, {Cheng}, \&
  {Zheng}}]{2011MNRAS.416.2174C}
{Cao}, X.-F., {Yu}, Y.-W., {Cheng}, K.~S., \& {Zheng}, X.-P. 2011, \mnras, 416,
  2174, \dodoi{10.1111/j.1365-2966.2011.19194.x}

\bibitem[{{Chary} {et~al.}(2007){Chary}, {Berger}, \&
  {Cowie}}]{2007ApJ...671..272C}
{Chary}, R., {Berger}, E., \& {Cowie}, L. 2007, \apj, 671, 272,
  \dodoi{10.1086/522692}

\bibitem[{{Chornock} {et~al.}(2014){Chornock}, {Berger}, {Fox}, {Fong},
  {Laskar}, \& {Roth}}]{2014arXiv1405.7400C}
{Chornock}, R., {Berger}, E., {Fox}, D.~B., {et~al.} 2014, arXiv e-prints,
  arXiv:1405.7400, \dodoi{10.48550/arXiv.1405.7400}

\bibitem[{{Chornock} {et~al.}(2013){Chornock}, {Berger}, {Fox}, {Lunnan},
  {Drout}, {Fong}, {Laskar}, \& {Roth}}]{2013ApJ...774...26C}
---. 2013, \apj, 774, 26, \dodoi{10.1088/0004-637X/774/1/26}

\bibitem[{{Ciardi} \& {Loeb}(2000)}]{2000ApJ...540..687C}
{Ciardi}, B., \& {Loeb}, A. 2000, \apj, 540, 687, \dodoi{10.1086/309384}

\bibitem[{{Ciardi} {et~al.}(2015){Ciardi}, {Inoue}, {Abdalla}, {Asad},
  {Bernardi}, {Bolton}, {Brentjens}, {de Bruyn}, {Chapman}, {Daiboo},
  {Fernandez}, {Ghosh}, {Graziani}, {Harker}, {Iliev}, {Jeli{\'c}}, {Jensen},
  {Kazemi}, {Koopmans}, {Martinez}, {Maselli}, {Mellema}, {Offringa}, {Pandey},
  {Schaye}, {Thomas}, {Vedantham}, {Yatawatta}, \&
  {Zaroubi}}]{2015MNRAS.453..101C}
{Ciardi}, B., {Inoue}, S., {Abdalla}, F.~B., {et~al.} 2015, \mnras, 453, 101,
  \dodoi{10.1093/mnras/stv1640}

\bibitem[{{Cucchiara} {et~al.}(2011){Cucchiara}, {Levan}, {Fox}, {Tanvir},
  {Ukwatta}, {Berger}, {Kr{\"u}hler}, {K{\"u}pc{\"u} Yolda{\c{s}}}, {Wu},
  {Toma}, {Greiner}, {Olivares}, {Rowlinson}, {Amati}, {Sakamoto}, {Roth},
  {Stephens}, {Fritz}, {Fynbo}, {Hjorth}, {Malesani}, {Jakobsson}, {Wiersema},
  {O'Brien}, {Soderberg}, {Foley}, {Fruchter}, {Rhoads}, {Rutledge}, {Schmidt},
  {Dopita}, {Podsiadlowski}, {Willingale}, {Wolf}, {Kulkarni}, \&
  {D'Avanzo}}]{2011ApJ...736....7C}
{Cucchiara}, A., {Levan}, A.~J., {Fox}, D.~B., {et~al.} 2011, Astrophys. J.,
  736, 7, \dodoi{10.1088/0004-637X/736/1/7}

\bibitem[{{Daigne} {et~al.}(2006){Daigne}, {Rossi}, \&
  {Mochkovitch}}]{2006MNRAS.372.1034D}
{Daigne}, F., {Rossi}, E.~M., \& {Mochkovitch}, R. 2006, \mnras, 372, 1034,
  \dodoi{10.1111/j.1365-2966.2006.10837.x}

\bibitem[{{de Souza} {et~al.}(2011){de Souza}, {Yoshida}, \&
  {Ioka}}]{2011A&A...533A..32D}
{de Souza}, R.~S., {Yoshida}, N., \& {Ioka}, K. 2011, \aap, 533, A32,
  \dodoi{10.1051/0004-6361/201117242}

\bibitem[{{Deng} {et~al.}(2016){Deng}, {Wang}, {Guo}, {Lu}, {Wang}, {Wei},
  {Wu}, \& {Liang}}]{2016ApJ...820...66D}
{Deng}, C.-M., {Wang}, X.-G., {Guo}, B.-B., {et~al.} 2016, \apj, 820, 66,
  \dodoi{10.3847/0004-637X/820/1/66}

\bibitem[{{Dichiara} {et~al.}(2023){Dichiara}, {Tsang}, {Troja}, {Neill},
  {Norris}, \& {Yang}}]{2023ApJ...954L..29D}
{Dichiara}, S., {Tsang}, D., {Troja}, E., {et~al.} 2023, \apjl, 954, L29,
  \dodoi{10.3847/2041-8213/acf21d}

\bibitem[{{Dong} {et~al.}(2022){Dong}, {Li}, {Zhang}, \&
  {Zhang}}]{2022MNRAS.513.1078D}
{Dong}, X.~F., {Li}, X.~J., {Zhang}, Z.~B., \& {Zhang}, X.~L. 2022, \mnras,
  513, 1078, \dodoi{10.1093/mnras/stac949}

\bibitem[{{Fausey} {et~al.}(2025{\natexlab{a}}){Fausey}, {van der Horst},
  {Tanvir}, {Wiersema}, {Fynbo}, {Hartmann}, \& {de Ugarte
  Postigo}}]{2025ApJ...985...28F}
{Fausey}, H.~M., {van der Horst}, A.~J., {Tanvir}, N.~R., {et~al.}
  2025{\natexlab{a}}, \apj, 985, 28, \dodoi{10.3847/1538-4357/adc5fc}

\bibitem[{{Fausey} {et~al.}(2025{\natexlab{b}}){Fausey}, {Vejlgaard}, {van der
  Horst}, {Heintz}, {Izzo}, {Malesani}, {Wiersema}, {Fynbo}, {Tanvir},
  {Vergani}, {Saccardi}, {Rossi}, {Campana}, {Covino}, {D'Elia}, {De Pasquale},
  {Hartmann}, {Jakobsson}, {Kouveliotou}, {Levan}, {Martin-Carrillo},
  {Melandri}, {Palmerio}, {Pugliese}, \& {Salvaterra}}]{2025MNRAS.536.2839F}
{Fausey}, H.~M., {Vejlgaard}, S., {van der Horst}, A.~J., {et~al.}
  2025{\natexlab{b}}, \mnras, 536, 2839, \dodoi{10.1093/mnras/stae2757}

\bibitem[{{Firmani} {et~al.}(2004){Firmani}, {Avila-Reese}, {Ghisellini}, \&
  {Tutukov}}]{2004ApJ...611.1033F}
{Firmani}, C., {Avila-Reese}, V., {Ghisellini}, G., \& {Tutukov}, A.~V. 2004,
  \apj, 611, 1033, \dodoi{10.1086/422186}

\bibitem[{{Gao} {et~al.}(2025){Gao}, {Geng}, {Liang}, {Sun}, {Xu}, {Wu},
  {Huang}, {Dai}, \& {Yuan}}]{2025ApJ...986..106G}
{Gao}, H.-X., {Geng}, J.-J., {Liang}, Y.-F., {et~al.} 2025, \apj, 986, 106,
  \dodoi{10.3847/1538-4357/adceb1}

\bibitem[{{Ghirlanda} \& {Salvaterra}(2022)}]{2022ApJ...932...10G}
{Ghirlanda}, G., \& {Salvaterra}, R. 2022, \apj, 932, 10,
  \dodoi{10.3847/1538-4357/ac6e43}

\bibitem[{{Ghirlanda} {et~al.}(2013){Ghirlanda}, {Ghisellini}, {Salvaterra},
  {Nava}, {Burlon}, {Tagliaferri}, {Campana}, {D'Avanzo}, \&
  {Melandri}}]{2013MNRAS.428.1410G}
{Ghirlanda}, G., {Ghisellini}, G., {Salvaterra}, R., {et~al.} 2013, \mnras,
  428, 1410, \dodoi{10.1093/mnras/sts128}

\bibitem[{{Ghirlanda} {et~al.}(2015){Ghirlanda}, {Salvaterra}, {Ghisellini},
  {Mereghetti}, {Tagliaferri}, {Campana}, {Osborne}, {O'Brien}, {Tanvir},
  {Willingale}, {Amati}, {Basa}, {Bernardini}, {Burlon}, {Covino}, {D'Avanzo},
  {Frontera}, {G{\"o}tz}, {Melandri}, {Nava}, {Piro}, \&
  {Vergani}}]{Ghirlanda2015MNRAS}
{Ghirlanda}, G., {Salvaterra}, R., {Ghisellini}, G., {et~al.} 2015, \mnras,
  448, 2514, \dodoi{10.1093/mnras/stv183}

\bibitem[{{Ghirlanda} {et~al.}(2021){Ghirlanda}, {Salvaterra}, {Toffano},
  {Ronchini}, {Guidorzi}, {Oganesyan}, {Ascenzi}, {Bernardini}, {Camisasca},
  {Mereghetti}, {Nava}, {Ravasio}, {Branchesi}, {Castro-Tirado}, {Amati},
  {Blain}, {Bozzo}, {O'Brien}, {G{\"o}tz}, {Le Floch}, {Osborne}, {Rosati},
  {Stratta}, {Tanvir}, {Bogomazov}, {D'Avanzo}, {Hafizi}, {Mandhai},
  {Melandri}, {Peer}, {Topinka}, {Vergani}, \& {Zane}}]{2021ExA....52..277G}
{Ghirlanda}, G., {Salvaterra}, R., {Toffano}, M., {et~al.} 2021, Experimental
  Astronomy, 52, 277, \dodoi{10.1007/s10686-021-09763-3}

\bibitem[{{Gou} {et~al.}(2004){Gou}, {M{\'e}sz{\'a}ros}, {Abel}, \&
  {Zhang}}]{2004ApJ...604..508G}
{Gou}, L.~J., {M{\'e}sz{\'a}ros}, P., {Abel}, T., \& {Zhang}, B. 2004, \apj,
  604, 508, \dodoi{10.1086/382061}

\bibitem[{{Greiner} {et~al.}(2009){Greiner}, {Kr{\"u}hler}, {Fynbo}, {Rossi},
  {Schwarz}, {Klose}, {Savaglio}, {Tanvir}, {McBreen}, {Totani}, {Zhang}, {Wu},
  {Watson}, {Barthelmy}, {Beardmore}, {Ferrero}, {Gehrels}, {Kann}, {Kawai},
  {Yolda{\c{s}}}, {M{\'e}sz{\'a}ros}, {Milvang-Jensen}, {Oates}, {Pierini},
  {Schady}, {Toma}, {Vreeswijk}, {Yolda{\c{s}}}, {Zhang}, {Afonso}, {Aoki},
  {Burrows}, {Clemens}, {Filgas}, {Haiman}, {Hartmann}, {Hasinger}, {Hjorth},
  {Jehin}, {Levan}, {Liang}, {Malesani}, {Pyo}, {Schulze}, {Szokoly}, {Terada},
  \& {Wiersema}}]{2009ApJ...693.1610G}
{Greiner}, J., {Kr{\"u}hler}, T., {Fynbo}, J.~P.~U., {et~al.} 2009, Astrophys.
  J., 693, 1610, \dodoi{10.1088/0004-637X/693/2/1610}

\bibitem[{{Gruber} {et~al.}(2014){Gruber}, {Goldstein}, {Weller von Ahlefeld},
  {Narayana Bhat}, {Bissaldi}, {Briggs}, {Byrne}, {Cleveland}, {Connaughton},
  {Diehl}, {Fishman}, {Fitzpatrick}, {Foley}, {Gibby}, {Giles}, {Greiner},
  {Guiriec}, {van der Horst}, {von Kienlin}, {Kouveliotou}, {Layden}, {Lin},
  {Meegan}, {McGlynn}, {Paciesas}, {Pelassa}, {Preece}, {Rau}, {Wilson-Hodge},
  {Xiong}, {Younes}, \& {Yu}}]{2014ApJS..211...12G}
{Gruber}, D., {Goldstein}, A., {Weller von Ahlefeld}, V., {et~al.} 2014, \apjs,
  211, 12, \dodoi{10.1088/0067-0049/211/1/12}

\bibitem[{{Guetta} \& {Piran}(2007)}]{2007JCAP...07..003G}
{Guetta}, D., \& {Piran}, T. 2007, \jcap, 2007, 003,
  \dodoi{10.1088/1475-7516/2007/07/003}

\bibitem[{{Guetta} {et~al.}(2005){Guetta}, {Piran}, \&
  {Waxman}}]{2005ApJ...619..412G}
{Guetta}, D., {Piran}, T., \& {Waxman}, E. 2005, \apj, 619, 412,
  \dodoi{10.1086/423125}

\bibitem[{{Hartoog} {et~al.}(2015){Hartoog}, {Malesani}, {Fynbo}, {Goto},
  {Kr{\"u}hler}, {Vreeswijk}, {De Cia}, {Xu}, {M{\o}ller}, {Covino}, {D'Elia},
  {Flores}, {Goldoni}, {Hjorth}, {Jakobsson}, {Krogager}, {Kaper}, {Ledoux},
  {Levan}, {Milvang-Jensen}, {Sollerman}, {Sparre}, {Tagliaferri}, {Tanvir},
  {de Ugarte Postigo}, {Vergani}, {Wiersema}, {Datson}, {Salinas}, {Mikkelsen},
  \& {Aghanim}}]{2015A&A...580A.139H}
{Hartoog}, O.~E., {Malesani}, D., {Fynbo}, J.~P.~U., {et~al.} 2015, \aap, 580,
  A139, \dodoi{10.1051/0004-6361/201425001}

\bibitem[{{Hopkins} \& {Beacom}(2006)}]{Hopkins2006ApJ}
{Hopkins}, A.~M., \& {Beacom}, J.~F. 2006, \apj, 651, 142,
  \dodoi{10.1086/506610}

\bibitem[{{Horv{\'a}th} \& {T{\'o}th}(2016)}]{2016Ap&SS.361..155H}
{Horv{\'a}th}, I., \& {T{\'o}th}, B.~G. 2016, \apss, 361, 155,
  \dodoi{10.1007/s10509-016-2748-6}

\bibitem[{{Ioka} \& {M{\'e}sz{\'a}ros}(2005)}]{2005ApJ...619..684I}
{Ioka}, K., \& {M{\'e}sz{\'a}ros}, P. 2005, \apj, 619, 684,
  \dodoi{10.1086/426785}

\bibitem[{{Jakobsson} {et~al.}(2006){Jakobsson}, {Levan}, {Fynbo}, {Priddey},
  {Hjorth}, {Tanvir}, {Watson}, {Jensen}, {Sollerman}, {Natarajan},
  {Gorosabel}, {Castro Cer{\'o}n}, {Pedersen}, {Pursimo}, {{\'A}rnad{\'o}ttir},
  {Castro-Tirado}, {Davis}, {Deeg}, {Fiuza}, {Mikolaitis}, \&
  {Sousa}}]{2006A&A...447..897J}
{Jakobsson}, P., {Levan}, A., {Fynbo}, J.~P.~U., {et~al.} 2006, \aap, 447, 897,
  \dodoi{10.1051/0004-6361:20054287}

\bibitem[{{Kaneko} {et~al.}(2006){Kaneko}, {Preece}, {Briggs}, {Paciesas},
  {Meegan}, \& {Band}}]{Kaneko2006ApJS}
{Kaneko}, Y., {Preece}, R.~D., {Briggs}, M.~S., {et~al.} 2006, \apjs, 166, 298,
  \dodoi{10.1086/505911}

\bibitem[{{Kawai} {et~al.}(2006){Kawai}, {Kosugi}, {Aoki}, {Yamada}, {Totani},
  {Ohta}, {Iye}, {Hattori}, {Aoki}, {Furusawa}, {Hurley}, {Kawabata},
  {Kobayashi}, {Komiyama}, {Mizumoto}, {Nomoto}, {Noumaru}, {Ogasawara},
  {Sato}, {Sekiguchi}, {Shirasaki}, {Suzuki}, {Takata}, {Tamagawa}, {Terada},
  {Watanabe}, {Yatsu}, \& {Yoshida}}]{2006Natur.440..184K}
{Kawai}, N., {Kosugi}, G., {Aoki}, K., {et~al.} 2006, Nature, 440, 184,
  \dodoi{10.1038/nature04498}

\bibitem[{{Kinugawa} {et~al.}(2019){Kinugawa}, {Harikane}, \&
  {Asano}}]{2019ApJ...878..128K}
{Kinugawa}, T., {Harikane}, Y., \& {Asano}, K. 2019, \apj, 878, 128,
  \dodoi{10.3847/1538-4357/ab2188}

\bibitem[{{Kistler} {et~al.}(2009){Kistler}, {Y{\"u}ksel}, {Beacom}, {Hopkins},
  \& {Wyithe}}]{2009ApJ...705L.104K}
{Kistler}, M.~D., {Y{\"u}ksel}, H., {Beacom}, J.~F., {Hopkins}, A.~M., \&
  {Wyithe}, J.~S.~B. 2009, \apjl, 705, L104,
  \dodoi{10.1088/0004-637X/705/2/L104}

\bibitem[{{Kistler} {et~al.}(2008){Kistler}, {Y{\"u}ksel}, {Beacom}, \&
  {Stanek}}]{2008ApJ...673L.119K}
{Kistler}, M.~D., {Y{\"u}ksel}, H., {Beacom}, J.~F., \& {Stanek}, K.~Z. 2008,
  \apjl, 673, L119, \dodoi{10.1086/527671}

\bibitem[{{Kouveliotou} {et~al.}(1993){Kouveliotou}, {Meegan}, {Fishman},
  {Bhat}, {Briggs}, {Koshut}, {Paciesas}, \& {Pendleton}}]{1993ApJ...413L.101K}
{Kouveliotou}, C., {Meegan}, C.~A., {Fishman}, G.~J., {et~al.} 1993, \apjl,
  413, L101, \dodoi{10.1086/186969}

\bibitem[{{Lan} {et~al.}(2021){Lan}, {Wei}, {Zeng}, {Li}, \&
  {Wu}}]{2021MNRAS.508...52L}
{Lan}, G.-X., {Wei}, J.-J., {Zeng}, H.-D., {Li}, Y., \& {Wu}, X.-F. 2021,
  \mnras, 508, 52, \dodoi{10.1093/mnras/stab2508}

\bibitem[{{Lan} {et~al.}(2019){Lan}, {Zeng}, {Wei}, \&
  {Wu}}]{2019MNRAS.488.4607L}
{Lan}, G.-X., {Zeng}, H.-D., {Wei}, J.-J., \& {Wu}, X.-F. 2019, \mnras, 488,
  4607, \dodoi{10.1093/mnras/stz2011}

\bibitem[{{Laskar} {et~al.}(2014){Laskar}, {Berger}, {Tanvir}, {Zauderer},
  {Margutti}, {Levan}, {Perley}, {Fong}, {Wiersema}, {Menten}, \&
  {Hrudkova}}]{2014ApJ...781....1L}
{Laskar}, T., {Berger}, E., {Tanvir}, N., {et~al.} 2014, Astrophys. J., 781, 1,
  \dodoi{10.1088/0004-637X/781/1/1}

\bibitem[{{Le} \& {Dermer}(2007)}]{2007ApJ...661..394L}
{Le}, T., \& {Dermer}, C.~D. 2007, \apj, 661, 394, \dodoi{10.1086/513460}

\bibitem[{{Le} {et~al.}(2020){Le}, {Ratke}, \& {Mehta}}]{2020MNRAS.493.1479L}
{Le}, T., {Ratke}, C., \& {Mehta}, V. 2020, \mnras, 493, 1479,
  \dodoi{10.1093/mnras/staa366}

\bibitem[{{Levan} {et~al.}(2024){Levan}, {Gompertz}, {Salafia}, {Bulla},
  {Burns}, {Hotokezaka}, {Izzo}, {Lamb}, {Malesani}, {Oates}, {Ravasio}, {Rouco
  Escorial}, {Schneider}, {Sarin}, {Schulze}, {Tanvir}, {Ackley}, {Anderson},
  {Brammer}, {Christensen}, {Dhillon}, {Evans}, {Fausnaugh}, {Fong},
  {Fruchter}, {Fryer}, {Fynbo}, {Gaspari}, {Heintz}, {Hjorth}, {Kennea},
  {Kennedy}, {Laskar}, {Leloudas}, {Mandel}, {Martin-Carrillo}, {Metzger},
  {Nicholl}, {Nugent}, {Palmerio}, {Pugliese}, {Rastinejad}, {Rhodes}, {Rossi},
  {Saccardi}, {Smartt}, {Stevance}, {Tohuvavohu}, {van der Horst}, {Vergani},
  {Watson}, {Barclay}, {Bhirombhakdi}, {Breedt}, {Breeveld}, {Brown},
  {Campana}, {Chrimes}, {D'Avanzo}, {D'Elia}, {De Pasquale}, {Dyer},
  {Galloway}, {Garbutt}, {Green}, {Hartmann}, {Jakobsson}, {Kerry},
  {Kouveliotou}, {Langeroodi}, {Le Floc'h}, {Leung}, {Littlefair}, {Munday},
  {O'Brien}, {Parsons}, {Pelisoli}, {Sahman}, {Salvaterra}, {Sbarufatti},
  {Steeghs}, {Tagliaferri}, {Th{\"o}ne}, {de Ugarte Postigo}, \&
  {Kann}}]{2024Natur.626..737L}
{Levan}, A.~J., {Gompertz}, B.~P., {Salafia}, O.~S., {et~al.} 2024, \nat, 626,
  737, \dodoi{10.1038/s41586-023-06759-1}

\bibitem[{{Li}(2008)}]{Li2008MNRAS}
{Li}, L.-X. 2008, \mnras, 388, 1487, \dodoi{10.1111/j.1365-2966.2008.13488.x}

\bibitem[{{Liang} {et~al.}(2007){Liang}, {Zhang}, {Virgili}, \&
  {Dai}}]{2007ApJ...662.1111L}
{Liang}, E., {Zhang}, B., {Virgili}, F., \& {Dai}, Z.~G. 2007, \apj, 662, 1111,
  \dodoi{10.1086/517959}

\bibitem[{{Lien} {et~al.}(2016){Lien}, {Sakamoto}, {Barthelmy}, {Baumgartner},
  {Cannizzo}, {Chen}, {Collins}, {Cummings}, {Gehrels}, {Krimm}, {Markwardt},
  {Palmer}, {Stamatikos}, {Troja}, \& {Ukwatta}}]{2016ApJ...829....7L}
{Lien}, A., {Sakamoto}, T., {Barthelmy}, S.~D., {et~al.} 2016, \apj, 829, 7,
  \dodoi{10.3847/0004-637X/829/1/7}

\bibitem[{{Littlejohns} {et~al.}(2013){Littlejohns}, {Tanvir}, {Willingale},
  {Evans}, {O'Brien}, \& {Levan}}]{2013MNRAS.436.3640L}
{Littlejohns}, O.~M., {Tanvir}, N.~R., {Willingale}, R., {et~al.} 2013, \mnras,
  436, 3640, \dodoi{10.1093/mnras/stt1841}

\bibitem[{{Liu} {et~al.}(2025){Liu}, {Sun}, {Xu}, {Svinkin}, {Delaunay},
  {Tanvir}, {Gao}, {Zhang}, {Chen}, {Wu}, {Zhang}, {Yuan}, {An}, {Bruni},
  {Frederiks}, {Ghirlanda}, {Hu}, {Li}, {Li}, {Li}, {Malesani}, {Piro},
  {Raman}, {Ricci}, {Troja}, {Vergani}, {Wu}, {Yang}, {Zhang}, {Zhu}, {de
  Ugarte Postigo}, {Demin}, {Dobie}, {Fan}, {Fu}, {Fynbo}, {Geng}, {Gianfagna},
  {Hu}, {Huang}, {Jiang}, {Jonker}, {Julakanti}, {Kennea}, {Kokomov},
  {Kuulkers}, {Lei}, {Leung}, {Levan}, {Li}, {Li}, {Littlefair}, {Liu},
  {Lysenko}, {Ma}, {Martin-Carrillo}, {O'Brien}, {Parsotan},
  {Quirola-V{\'a}squez}, {Ridnaia}, {Ronchini}, {Rossi}, {Mata-S{\'a}nchez},
  {Schneider}, {Shen}, {Thakur}, {Tohuvavohu}, {Torres}, {Tsvetkova}, {Ulanov},
  {Wei}, {Xiao}, {Yin}, {Bai}, {Burwitz}, {Cai}, {Chen}, {Chen}, {Chen},
  {Chen}, {Chen}, {Chen}, {Cheng}, {Cordier}, {Cui}, {Cui}, {Dai}, {Dai},
  {Eder}, {Eyles-Ferris}, {Fan}, {Feldman}, {Feng}, {Feng}, {Friedrich}, {Gao},
  {Gonzalez}, {Guan}, {Han}, {Han}, {Hou}, {Hu}, {Hu}, {Huang}, {Huo},
  {Hutchinson}, {Ji}, {Jia}, {Jia}, {Jiang}, {Jin}, {Jin}, {Jin}, {Keereman},
  {Lerman}, {Li}, {Li}, {Li}, {Li}, {Li}, {Lian}, {Liang}, {Ling}, {Liu},
  {Liu}, {Liu}, {Liu}, {Liu}, {Lu}, {L{\"u}}, {Luo}, {Ma}, {Ma}, {Mao}, {Mao},
  {McHugh}, {Meidinger}, {Nandra}, {Osborne}, {Pan}, {Pan}, {Ravasio}, {Rau},
  {Rea}, {Rehman}, {Sanders}, {Santovincenzo}, {Song}, {Su}, {Sun}, {Sun},
  {Sun}, {Tan}, {Tang}, {Tao}, {Tong}, {Wang}, {Wang}, {Wang}, {Wang}, {Wang},
  {Wang}, {Wang}, {Wang}, {Wang}, {Wei}, {Willingale}, {Xiong}, {Xu}, {Xu},
  {Xu}, {Xu}, {Xu}, {Xue}, {Xue}, {Yan}, {Yang}, {Yang}, {Yang}, {Yang}, {Yu},
  {Zhang}, {Zhang}, {Zhang}, {Zhang}, {Zhang}, {Zhang}, {Zhang}, {Zhang},
  {Zhang}, {Zhao}, {Zhao}, {Zhao}, {Zhao}, {Zhou}, {Zhou}, {Zhu}, {Zhu}, \&
  {Zuo}}]{2025NatAs...9..564L}
{Liu}, Y., {Sun}, H., {Xu}, D., {et~al.} 2025, Nature Astronomy, 9, 564,
  \dodoi{10.1038/s41550-024-02449-8}

\bibitem[{{Llamas Lanza} {et~al.}(2024){Llamas Lanza}, {Godet}, {Arcier},
  {Yassine}, {Atteia}, \& {Bouchet}}]{2024A&A...685A.163L}
{Llamas Lanza}, M., {Godet}, O., {Arcier}, B., {et~al.} 2024, \aap, 685, A163,
  \dodoi{10.1051/0004-6361/202347966}

\bibitem[{{Lloyd-Ronning} {et~al.}(2019){Lloyd-Ronning}, {Aykutalp}, \&
  {Johnson}}]{2019MNRAS.488.5823L}
{Lloyd-Ronning}, N.~M., {Aykutalp}, A., \& {Johnson}, J.~L. 2019, \mnras, 488,
  5823, \dodoi{10.1093/mnras/stz2155}

\bibitem[{{Lloyd-Ronning} {et~al.}(2002){Lloyd-Ronning}, {Fryer}, \&
  {Ramirez-Ruiz}}]{2002ApJ...574..554L}
{Lloyd-Ronning}, N.~M., {Fryer}, C.~L., \& {Ramirez-Ruiz}, E. 2002, \apj, 574,
  554, \dodoi{10.1086/341059}

\bibitem[{{Lu} {et~al.}(2012){Lu}, {Wei}, {Qin}, \&
  {Liang}}]{2012ApJ...745..168L}
{Lu}, R.-J., {Wei}, J.-J., {Qin}, S.-F., \& {Liang}, E.-W. 2012, \apj, 745,
  168, \dodoi{10.1088/0004-637X/745/2/168}

\bibitem[{{Madau} \& {Dickinson}(2014)}]{2014ARA&A..52..415M}
{Madau}, P., \& {Dickinson}, M. 2014, \araa, 52, 415,
  \dodoi{10.1146/annurev-astro-081811-125615}

\bibitem[{{Malesani} {et~al.}(2025){Malesani}, {Pugliese}, {Fynbo},
  {Schneider}, {D'Elia}, {de Ugarte Postigo}, {Izzo}, {Jonker}, {Levan},
  {Palmerio}, {Rakotondrainibe}, {Saccardi}, {Tanvir}, {Thakur}, {Vergani},
  {Xu}, {Zhu}, \& {Stargate Collaboration}}]{2025GCN.39732....1M}
{Malesani}, D.~B., {Pugliese}, G., {Fynbo}, J.~P.~U., {et~al.} 2025, GRB
  Coordinates Network, 39732, 1

\bibitem[{{Matsumoto} {et~al.}(2024){Matsumoto}, {Harikane}, {Maeda}, \&
  {Ioka}}]{2024ApJ...976L..16M}
{Matsumoto}, T., {Harikane}, Y., {Maeda}, K., \& {Ioka}, K. 2024, \apjl, 976,
  L16, \dodoi{10.3847/2041-8213/ad8ce0}

\bibitem[{{Matsumoto} {et~al.}(2015){Matsumoto}, {Nakauchi}, {Ioka}, {Heger},
  \& {Nakamura}}]{2015ApJ...810...64M}
{Matsumoto}, T., {Nakauchi}, D., {Ioka}, K., {Heger}, A., \& {Nakamura}, T.
  2015, \apj, 810, 64, \dodoi{10.1088/0004-637X/810/1/64}

\bibitem[{{Matsumoto} {et~al.}(2016){Matsumoto}, {Nakauchi}, {Ioka}, \&
  {Nakamura}}]{2016ApJ...823...83M}
{Matsumoto}, T., {Nakauchi}, D., {Ioka}, K., \& {Nakamura}, T. 2016, \apj, 823,
  83, \dodoi{10.3847/0004-637X/823/2/83}

\bibitem[{{Melandri} {et~al.}(2015){Melandri}, {Bernardini}, {D'Avanzo},
  {S{\'a}nchez-Ram{\'\i}rez}, {Nappo}, {Nava}, {Japelj}, {de Ugarte Postigo},
  {Oates}, {Campana}, {Covino}, {D'Elia}, {Ghirlanda}, {Gafton}, {Ghisellini},
  {Gnedin}, {Goldoni}, {Gorosabel}, {Libbrecht}, {Malesani}, {Salvaterra},
  {Th{\"o}ne}, {Vergani}, {Xu}, \& {Tagliaferri}}]{2015A&A...581A..86M}
{Melandri}, A., {Bernardini}, M.~G., {D'Avanzo}, P., {et~al.} 2015, Astron.
  Astrophys., 581, A86, \dodoi{10.1051/0004-6361/201526660}

\bibitem[{{M{\'e}sz{\'a}ros} \& {Rees}(2010)}]{2010ApJ...715..967M}
{M{\'e}sz{\'a}ros}, P., \& {Rees}, M.~J. 2010, \apj, 715, 967,
  \dodoi{10.1088/0004-637X/715/2/967}

\bibitem[{{Narayana Bhat} {et~al.}(2016){Narayana Bhat}, {Meegan}, {von
  Kienlin}, {Paciesas}, {Briggs}, {Burgess}, {Burns}, {Chaplin}, {Cleveland},
  {Collazzi}, {Connaughton}, {Diekmann}, {Fitzpatrick}, {Gibby}, {Giles},
  {Goldstein}, {Greiner}, {Jenke}, {Kippen}, {Kouveliotou}, {Mailyan},
  {McBreen}, {Pelassa}, {Preece}, {Roberts}, {Sparke}, {Stanbro}, {Veres},
  {Wilson-Hodge}, {Xiong}, {Younes}, {Yu}, \& {Zhang}}]{2016ApJS..223...28N}
{Narayana Bhat}, P., {Meegan}, C.~A., {von Kienlin}, A., {et~al.} 2016, \apjs,
  223, 28, \dodoi{10.3847/0067-0049/223/2/28}

\bibitem[{{Natarajan} {et~al.}(2005){Natarajan}, {Albanna}, {Hjorth},
  {Ramirez-Ruiz}, {Tanvir}, \& {Wijers}}]{2005MNRAS.364L...8N}
{Natarajan}, P., {Albanna}, B., {Hjorth}, J., {et~al.} 2005, \mnras, 364, L8,
  \dodoi{10.1111/j.1745-3933.2005.00094.x}

\bibitem[{{Nava} {et~al.}(2011){Nava}, {Ghirlanda}, {Ghisellini}, \&
  {Celotti}}]{Nava2011A&A}
{Nava}, L., {Ghirlanda}, G., {Ghisellini}, G., \& {Celotti}, A. 2011, \aap,
  530, A21, \dodoi{10.1051/0004-6361/201016270}

\bibitem[{{Nava} {et~al.}(2012){Nava}, {Salvaterra}, {Ghirlanda}, {Ghisellini},
  {Campana}, {Covino}, {Cusumano}, {D'Avanzo}, {D'Elia}, {Fugazza}, {Melandri},
  {Sbarufatti}, {Vergani}, \& {Tagliaferri}}]{Nava2012MNRAS}
{Nava}, L., {Salvaterra}, R., {Ghirlanda}, G., {et~al.} 2012, \mnras, 421,
  1256, \dodoi{10.1111/j.1365-2966.2011.20394.x}

\bibitem[{{Paczy{\'n}ski}(1998)}]{1998ApJ...494L..45P}
{Paczy{\'n}ski}, B. 1998, \apjl, 494, L45, \dodoi{10.1086/311148}

\bibitem[{{Palmerio} \& {Daigne}(2021)}]{2021A&A...649A.166P}
{Palmerio}, J.~T., \& {Daigne}, F. 2021, \aap, 649, A166,
  \dodoi{10.1051/0004-6361/202039929}

\bibitem[{{Paul}(2018)}]{2018MNRAS.473.3385P}
{Paul}, D. 2018, \mnras, 473, 3385, \dodoi{10.1093/mnras/stx2511}

\bibitem[{{Perley} {et~al.}(2016){Perley}, {Kr{\"u}hler}, {Schulze}, {de Ugarte
  Postigo}, {Hjorth}, {Berger}, {Cenko}, {Chary}, {Cucchiara}, {Ellis}, {Fong},
  {Fynbo}, {Gorosabel}, {Greiner}, {Jakobsson}, {Kim}, {Laskar}, {Levan},
  {Micha{\l}owski}, {Milvang-Jensen}, {Tanvir}, {Th{\"o}ne}, \&
  {Wiersema}}]{2016ApJ...817....7P}
{Perley}, D.~A., {Kr{\"u}hler}, T., {Schulze}, S., {et~al.} 2016, \apj, 817, 7,
  \dodoi{10.3847/0004-637X/817/1/7}

\bibitem[{{Pescalli} {et~al.}(2015){Pescalli}, {Ghirlanda}, {Salafia},
  {Ghisellini}, {Nappo}, \& {Salvaterra}}]{2015MNRAS.447.1911P}
{Pescalli}, A., {Ghirlanda}, G., {Salafia}, O.~S., {et~al.} 2015, \mnras, 447,
  1911, \dodoi{10.1093/mnras/stu2482}

\bibitem[{{Pescalli} {et~al.}(2016){Pescalli}, {Ghirlanda}, {Salvaterra},
  {Ghisellini}, {Vergani}, {Nappo}, {Salafia}, {Melandri}, {Covino}, \&
  {G{\"o}tz}}]{2016A&A...587A..40P}
{Pescalli}, A., {Ghirlanda}, G., {Salvaterra}, R., {et~al.} 2016, \aap, 587,
  A40, \dodoi{10.1051/0004-6361/201526760}

\bibitem[{{Petrosian} {et~al.}(2015){Petrosian}, {Kitanidis}, \&
  {Kocevski}}]{2015ApJ...806...44P}
{Petrosian}, V., {Kitanidis}, E., \& {Kocevski}, D. 2015, \apj, 806, 44,
  \dodoi{10.1088/0004-637X/806/1/44}

\bibitem[{{Planck Collaboration} {et~al.}(2020)}]{2020AA...641A...6P}
{Planck Collaboration}, {et~al.} 2020, \aap, 641, A6,
  \dodoi{10.1051/0004-6361/201833910}

\bibitem[{{Porciani} \& {Madau}(2001)}]{2001ApJ...548..522P}
{Porciani}, C., \& {Madau}, P. 2001, \apj, 548, 522, \dodoi{10.1086/319027}

\bibitem[{{Qin} {et~al.}(2010){Qin}, {Liang}, {Lu}, {Wei}, \&
  {Zhang}}]{2010MNRAS.406..558Q}
{Qin}, S.-F., {Liang}, E.-W., {Lu}, R.-J., {Wei}, J.-Y., \& {Zhang}, S.-N.
  2010, \mnras, 406, 558, \dodoi{10.1111/j.1365-2966.2010.16691.x}

\bibitem[{{Qu} {et~al.}(2025){Qu}, {Man}, {Yang}, {Yi}, {Du}, \&
  {Wang}}]{2025ApJ...982..148Q}
{Qu}, Y.-K., {Man}, Z.-X., {Yang}, Y.-P., {et~al.} 2025, \apj, 982, 148,
  \dodoi{10.3847/1538-4357/adb849}

\bibitem[{{Qu} {et~al.}(2024){Qu}, {Man}, {Yi}, \&
  {Yang}}]{2024ApJ...976..170Q}
{Qu}, Y.-K., {Man}, Z.-X., {Yi}, S.-X., \& {Yang}, Y.-P. 2024, \apj, 976, 170,
  \dodoi{10.3847/1538-4357/ad88e7}

\bibitem[{{Rastinejad} {et~al.}(2022){Rastinejad}, {Gompertz}, {Levan}, {Fong},
  {Nicholl}, {Lamb}, {Malesani}, {Nugent}, {Oates}, {Tanvir}, {de Ugarte
  Postigo}, {Kilpatrick}, {Moore}, {Metzger}, {Ravasio}, {Rossi}, {Schroeder},
  {Jencson}, {Sand}, {Smith}, {Ag{\"u}{\'\i} Fern{\'a}ndez}, {Berger},
  {Blanchard}, {Chornock}, {Cobb}, {De Pasquale}, {Fynbo}, {Izzo}, {Kann},
  {Laskar}, {Marini}, {Paterson}, {Escorial}, {Sears}, \&
  {Th{\"o}ne}}]{2022Natur.612..223R}
{Rastinejad}, J.~C., {Gompertz}, B.~P., {Levan}, A.~J., {et~al.} 2022, \nat,
  612, 223, \dodoi{10.1038/s41586-022-05390-w}

\bibitem[{{Robertson} \& {Ellis}(2012)}]{2012ApJ...744...95R}
{Robertson}, B.~E., \& {Ellis}, R.~S. 2012, \apj, 744, 95,
  \dodoi{10.1088/0004-637X/744/2/95}

\bibitem[{{Rossi} {et~al.}(2022){Rossi}, {Frederiks}, {Kann}, {De Pasquale},
  {Pian}, {Lamb}, {D'Avanzo}, {Izzo}, {Levan}, {Malesani}, {Melandri}, {Nicuesa
  Guelbenzu}, {Schulze}, {Strausbaugh}, {Tanvir}, {Amati}, {Campana},
  {Cucchiara}, {Ghirlanda}, {Della Valle}, {Klose}, {Salvaterra}, {Starling},
  {Stratta}, {Tsvetkova}, {Vergani}, {D'A{\`\i}}, {Burgarella}, {Covino},
  {D'Elia}, {de Ugarte Postigo}, {Fausey}, {Fynbo}, {Frontera}, {Guidorzi},
  {Heintz}, {Masetti}, {Maiorano}, {Mundell}, {Oates}, {Page}, {Palazzi},
  {Palmerio}, {Pugliese}, {Rau}, {Saccardi}, {Sbarufatti}, {Svinkin},
  {Tagliaferri}, {van der Horst}, {Watson}, {Ulanov}, {Wiersema}, {Xu}, \&
  {Zhang}}]{2022A&A...665A.125R}
{Rossi}, A., {Frederiks}, D.~D., {Kann}, D.~A., {et~al.} 2022, \aap, 665, A125,
  \dodoi{10.1051/0004-6361/202243225}

\bibitem[{{Saccardi} {et~al.}(2023){Saccardi}, {Vergani}, {De Cia}, {D'Elia},
  {Heintz}, {Izzo}, {Palmerio}, {Petitjean}, {Rossi}, {de Ugarte Postigo},
  {Christensen}, {Konstantopoulou}, {Levan}, {Malesani}, {M{\o}ller},
  {Ramburuth-Hurt}, {Salvaterra}, {Tanvir}, {Th{\"o}ne}, {Vejlgaard}, {Fynbo},
  {Kann}, {Schady}, {Watson}, {Wiersema}, {Campana}, {Covino}, {De Pasquale},
  {Fausey}, {Hartmann}, {van der Horst}, {Jakobsson}, {Palazzi}, {Pugliese},
  {Savaglio}, {Starling}, {Stratta}, \& {Zafar}}]{2023A&A...671A..84S}
{Saccardi}, A., {Vergani}, S.~D., {De Cia}, A., {et~al.} 2023, Astron.
  Astrophys., 671, A84, \dodoi{10.1051/0004-6361/202244205}

\bibitem[{{Saccardi} {et~al.}(2024){Saccardi}, {Malesani}, {Palmerio},
  {Vergani}, {Le Floc'h}, {Izzo}, {Levan}, {Fynbo}, {D'Avanzo}, {Rossi}, {de
  Ugarte Postigo}, \& {Stargate Collaboration}}]{2024GCN.35756....1S}
{Saccardi}, A., {Malesani}, D.~B., {Palmerio}, J.~T., {et~al.} 2024, GRB
  Coordinates Network, 35756, 1

\bibitem[{{Sakamoto} {et~al.}(2008){Sakamoto}, {Hullinger}, {Sato}, {Yamazaki},
  {Barbier}, {Barthelmy}, {Cummings}, {Fenimore}, {Gehrels}, {Krimm}, {Lamb},
  {Markwardt}, {Osborne}, {Palmer}, {Parsons}, {Stamatikos}, \&
  {Tueller}}]{2008ApJ...679..570S}
{Sakamoto}, T., {Hullinger}, D., {Sato}, G., {et~al.} 2008, \apj, 679, 570,
  \dodoi{10.1086/586884}

\bibitem[{{Salafia} {et~al.}(2016){Salafia}, {Ghisellini}, {Pescalli},
  {Ghirlanda}, \& {Nappo}}]{2016MNRAS.461.3607S}
{Salafia}, O.~S., {Ghisellini}, G., {Pescalli}, A., {Ghirlanda}, G., \&
  {Nappo}, F. 2016, \mnras, 461, 3607, \dodoi{10.1093/mnras/stw1549}

\bibitem[{{Salvaterra}(2015)}]{2015JHEAp...7...35S}
{Salvaterra}, R. 2015, Journal of High Energy Astrophysics, 7, 35,
  \dodoi{10.1016/j.jheap.2015.03.001}

\bibitem[{{Salvaterra} \& {Chincarini}(2007)}]{2007ApJ...656L..49S}
{Salvaterra}, R., \& {Chincarini}, G. 2007, \apjl, 656, L49,
  \dodoi{10.1086/512606}

\bibitem[{{Salvaterra} {et~al.}(2009{\natexlab{a}}){Salvaterra}, {Guidorzi},
  {Campana}, {Chincarini}, \& {Tagliaferri}}]{2009MNRAS.396..299S}
{Salvaterra}, R., {Guidorzi}, C., {Campana}, S., {Chincarini}, G., \&
  {Tagliaferri}, G. 2009{\natexlab{a}}, \mnras, 396, 299,
  \dodoi{10.1111/j.1365-2966.2008.14343.x}

\bibitem[{{Salvaterra} {et~al.}(2009{\natexlab{b}}){Salvaterra}, {Della Valle},
  {Campana}, {Chincarini}, {Covino}, {D'Avanzo}, {Fern{\'a}ndez-Soto},
  {Guidorzi}, {Mannucci}, {Margutti}, {Th{\"o}ne}, {Antonelli}, {Barthelmy},
  {de Pasquale}, {D'Elia}, {Fiore}, {Fugazza}, {Hunt}, {Maiorano}, {Marinoni},
  {Marshall}, {Molinari}, {Nousek}, {Pian}, {Racusin}, {Stella}, {Amati},
  {Andreuzzi}, {Cusumano}, {Fenimore}, {Ferrero}, {Giommi}, {Guetta},
  {Holland}, {Hurley}, {Israel}, {Mao}, {Markwardt}, {Masetti}, {Pagani},
  {Palazzi}, {Palmer}, {Piranomonte}, {Tagliaferri}, \&
  {Testa}}]{2009Natur.461.1258S}
{Salvaterra}, R., {Della Valle}, M., {Campana}, S., {et~al.}
  2009{\natexlab{b}}, Nature, 461, 1258, \dodoi{10.1038/nature08445}

\bibitem[{{Salvaterra} {et~al.}(2012){Salvaterra}, {Campana}, {Vergani},
  {Covino}, {D'Avanzo}, {Fugazza}, {Ghirlanda}, {Ghisellini}, {Melandri},
  {Nava}, {Sbarufatti}, {Flores}, {Piranomonte}, \&
  {Tagliaferri}}]{Salvaterra2012ApJ}
{Salvaterra}, R., {Campana}, S., {Vergani}, S.~D., {et~al.} 2012, \apj, 749,
  68, \dodoi{10.1088/0004-637X/749/1/68}

\bibitem[{{Suwa} \& {Ioka}(2011)}]{2011ApJ...726..107S}
{Suwa}, Y., \& {Ioka}, K. 2011, \apj, 726, 107,
  \dodoi{10.1088/0004-637X/726/2/107}

\bibitem[{{Tan} {et~al.}(2013){Tan}, {Cao}, \& {Yu}}]{2013ApJ...772L...8T}
{Tan}, W.-W., {Cao}, X.-F., \& {Yu}, Y.-W. 2013, \apjl, 772, L8,
  \dodoi{10.1088/2041-8205/772/1/L8}

\bibitem[{{Tan} \& {Wang}(2015)}]{2015MNRAS.454.1785T}
{Tan}, W.-W., \& {Wang}, F.~Y. 2015, \mnras, 454, 1785,
  \dodoi{10.1093/mnras/stv2007}

\bibitem[{{Tanvir} {et~al.}(2009){Tanvir}, {Fox}, {Levan}, {Berger},
  {Wiersema}, {Fynbo}, {Cucchiara}, {Kr{\"u}hler}, {Gehrels}, {Bloom},
  {Greiner}, {Evans}, {Rol}, {Olivares}, {Hjorth}, {Jakobsson}, {Farihi},
  {Willingale}, {Starling}, {Cenko}, {Perley}, {Maund}, {Duke}, {Wijers},
  {Adamson}, {Allan}, {Bremer}, {Burrows}, {Castro-Tirado}, {Cavanagh}, {de
  Ugarte Postigo}, {Dopita}, {Fatkhullin}, {Fruchter}, {Foley}, {Gorosabel},
  {Kennea}, {Kerr}, {Klose}, {Krimm}, {Komarova}, {Kulkarni}, {Moskvitin},
  {Mundell}, {Naylor}, {Page}, {Penprase}, {Perri}, {Podsiadlowski}, {Roth},
  {Rutledge}, {Sakamoto}, {Schady}, {Schmidt}, {Soderberg}, {Sollerman},
  {Stephens}, {Stratta}, {Ukwatta}, {Watson}, {Westra}, {Wold}, \&
  {Wolf}}]{2009Natur.461.1254T}
{Tanvir}, N.~R., {Fox}, D.~B., {Levan}, A.~J., {et~al.} 2009, Nature, 461,
  1254, \dodoi{10.1038/nature08459}

\bibitem[{{Tanvir} {et~al.}(2018){Tanvir}, {Laskar}, {Levan}, {Perley}, {Zabl},
  {Fynbo}, {Rhoads}, {Cenko}, {Greiner}, {Wiersema}, {Hjorth}, {Cucchiara},
  {Berger}, {Bremer}, {Cano}, {Cobb}, {Covino}, {D'Elia}, {Fong}, {Fruchter},
  {Goldoni}, {Hammer}, {Heintz}, {Jakobsson}, {Kann}, {Kaper}, {Klose},
  {Knust}, {Kr{\"u}hler}, {Malesani}, {Misra}, {Nicuesa Guelbenzu}, {Pugliese},
  {S{\'a}nchez-Ram{\'\i}rez}, {Schulze}, {Stanway}, {de Ugarte Postigo},
  {Watson}, {Wijers}, \& {Xu}}]{2018ApJ...865..107T}
{Tanvir}, N.~R., {Laskar}, T., {Levan}, A.~J., {et~al.} 2018, \apj, 865, 107,
  \dodoi{10.3847/1538-4357/aadba9}

\bibitem[{{Tanvir} {et~al.}(2019){Tanvir}, {Fynbo}, {de Ugarte Postigo},
  {Japelj}, {Wiersema}, {Malesani}, {Perley}, {Levan}, {Selsing}, {Cenko},
  {Kann}, {Milvang-Jensen}, {Berger}, {Cano}, {Chornock}, {Covino},
  {Cucchiara}, {D'Elia}, {Gargiulo}, {Goldoni}, {Gomboc}, {Heintz}, {Hjorth},
  {Izzo}, {Jakobsson}, {Kaper}, {Kr{\"u}hler}, {Laskar}, {Myers},
  {Piranomonte}, {Pugliese}, {Rossi}, {S{\'a}nchez-Ram{\'\i}rez}, {Schulze},
  {Sparre}, {Stanway}, {Tagliaferri}, {Th{\"o}ne}, {Vergani}, {Vreeswijk},
  {Wijers}, {Watson}, \& {Xu}}]{2019MNRAS.483.5380T}
{Tanvir}, N.~R., {Fynbo}, J.~P.~U., {de Ugarte Postigo}, A., {et~al.} 2019,
  \mnras, 483, 5380, \dodoi{10.1093/mnras/sty3460}

\bibitem[{{Toma} {et~al.}(2011){Toma}, {Sakamoto}, \&
  {M{\'e}sz{\'a}ros}}]{2011ApJ...731..127T}
{Toma}, K., {Sakamoto}, T., \& {M{\'e}sz{\'a}ros}, P. 2011, \apj, 731, 127,
  \dodoi{10.1088/0004-637X/731/2/127}

\bibitem[{{Totani}(1997)}]{1997ApJ...486L..71T}
{Totani}, T. 1997, \apjl, 486, L71, \dodoi{10.1086/310853}

\bibitem[{{Totani} {et~al.}(2006){Totani}, {Kawai}, {Kosugi}, {Aoki}, {Yamada},
  {Iye}, {Ohta}, \& {Hattori}}]{2006PASJ...58..485T}
{Totani}, T., {Kawai}, N., {Kosugi}, G., {et~al.} 2006, \pasj, 58, 485,
  \dodoi{10.1093/pasj/58.3.485}

\bibitem[{{Totani} {et~al.}(2014){Totani}, {Aoki}, {Hattori}, {Kosugi},
  {Niino}, {Hashimoto}, {Kawai}, {Ohta}, {Sakamoto}, \&
  {Yamada}}]{2014PASJ...66...63T}
{Totani}, T., {Aoki}, K., {Hattori}, T., {et~al.} 2014, \pasj, 66, 63,
  \dodoi{10.1093/pasj/psu032}

\bibitem[{{Troja} {et~al.}(2022){Troja}, {Fryer}, {O'Connor}, {Ryan},
  {Dichiara}, {Kumar}, {Ito}, {Gupta}, {Wollaeger}, {Norris}, {Kawai},
  {Butler}, {Aryan}, {Misra}, {Hosokawa}, {Murata}, {Niwano}, {Pandey},
  {Kutyrev}, {van Eerten}, {Chase}, {Hu}, {Caballero-Garcia}, \&
  {Castro-Tirado}}]{2022Natur.612..228T}
{Troja}, E., {Fryer}, C.~L., {O'Connor}, B., {et~al.} 2022, \nat, 612, 228,
  \dodoi{10.1038/s41586-022-05327-3}

\bibitem[{{Veres} {et~al.}(2024){Veres}, {Meegan}, \& {Fermi Gamma-ray Burst
  Monitor Team}}]{2024GCN.35755....1V}
{Veres}, P., {Meegan}, C., \& {Fermi Gamma-ray Burst Monitor Team}. 2024, GRB
  Coordinates Network, 35755, 1

\bibitem[{{Virgili} {et~al.}(2011){Virgili}, {Zhang}, {Nagamine}, \&
  {Choi}}]{2011MNRAS.417.3025V}
{Virgili}, F.~J., {Zhang}, B., {Nagamine}, K., \& {Choi}, J.-H. 2011, \mnras,
  417, 3025, \dodoi{10.1111/j.1365-2966.2011.19459.x}

\bibitem[{{von Kienlin} {et~al.}(2014){von Kienlin}, {Meegan}, {Paciesas},
  {Bhat}, {Bissaldi}, {Briggs}, {Burgess}, {Byrne}, {Chaplin}, {Cleveland},
  {Connaughton}, {Collazzi}, {Fitzpatrick}, {Foley}, {Gibby}, {Giles},
  {Goldstein}, {Greiner}, {Gruber}, {Guiriec}, {van der Horst}, {Kouveliotou},
  {Layden}, {McBreen}, {McGlynn}, {Pelassa}, {Preece}, {Rau}, {Tierney},
  {Wilson-Hodge}, {Xiong}, {Younes}, \& {Yu}}]{2014ApJS..211...13V}
{von Kienlin}, A., {Meegan}, C.~A., {Paciesas}, W.~S., {et~al.} 2014, \apjs,
  211, 13, \dodoi{10.1088/0067-0049/211/1/13}

\bibitem[{{von Kienlin} {et~al.}(2020){von Kienlin}, {Meegan}, {Paciesas},
  {Bhat}, {Bissaldi}, {Briggs}, {Burns}, {Cleveland}, {Gibby}, {Giles},
  {Goldstein}, {Hamburg}, {Hui}, {Kocevski}, {Mailyan}, {Malacaria},
  {Poolakkil}, {Preece}, {Roberts}, {Veres}, \&
  {Wilson-Hodge}}]{2020ApJ...893...46V}
---. 2020, \apj, 893, 46, \dodoi{10.3847/1538-4357/ab7a18}

\bibitem[{{{\v{S}}oltinsk{\'y}} {et~al.}(2025){{\v{S}}oltinsk{\'y}},
  {Kulkarni}, {Tendulkar}, \& {Bolton}}]{2025MNRAS.537..364S}
{{\v{S}}oltinsk{\'y}}, T., {Kulkarni}, G., {Tendulkar}, S.~P., \& {Bolton},
  J.~S. 2025, \mnras, 537, 364, \dodoi{10.1093/mnras/staf026}

\bibitem[{{Wanderman} \& {Piran}(2010)}]{2010MNRAS.406.1944W}
{Wanderman}, D., \& {Piran}, T. 2010, \mnras, 406, 1944,
  \dodoi{10.1111/j.1365-2966.2010.16787.x}

\bibitem[{{Wang}(2013)}]{2013A&A...556A..90W}
{Wang}, F.~Y. 2013, \aap, 556, A90, \dodoi{10.1051/0004-6361/201321623}

\bibitem[{{Wang} {et~al.}(2012){Wang}, {Bromm}, {Greif}, {Stacy}, {Dai},
  {Loeb}, \& {Cheng}}]{2012ApJ...760...27W}
{Wang}, F.~Y., {Bromm}, V., {Greif}, T.~H., {et~al.} 2012, \apj, 760, 27,
  \dodoi{10.1088/0004-637X/760/1/27}

\bibitem[{{Wang} \& {Dai}(2009)}]{2009MNRAS.400L..10W}
{Wang}, F.~Y., \& {Dai}, Z.~G. 2009, \mnras, 400, L10,
  \dodoi{10.1111/j.1745-3933.2009.00751.x}

\bibitem[{{Wang} {et~al.}(2015){Wang}, {Dai}, \& {Liang}}]{2015NewAR..67....1W}
{Wang}, F.~Y., {Dai}, Z.~G., \& {Liang}, E.~W. 2015, NewAR, 67, 1,
  \dodoi{10.1016/j.newar.2015.03.001}

\bibitem[{{Wei} {et~al.}(2016){Wei}, {Cordier}, {Antier}, {Antilogus},
  {Atteia}, {Bajat}, {Basa}, {Beckmann}, {Bernardini}, {Boissier}, {Bouchet},
  {Burwitz}, {Claret}, {Dai}, {Daigne}, {Deng}, {Dornic}, {Feng}, {Foglizzo},
  {Gao}, {Gehrels}, {Godet}, {Goldwurm}, {Gonzalez}, {Gosset}, {G{\"o}tz},
  {Gouiffes}, {Grise}, {Gros}, {Guilet}, {Han}, {Huang}, {Huang}, {Jouret},
  {Klotz}, {La Marle}, {Lachaud}, {Le Floch}, {Lee}, {Leroy}, {Li}, {Li}, {Li},
  {Liang}, {Lyu}, {Mercier}, {Migliori}, {Mochkovitch}, {O'Brien}, {Osborne},
  {Paul}, {Perinati}, {Petitjean}, {Piron}, {Qiu}, {Rau}, {Rodriguez},
  {Schanne}, {Tanvir}, {Vangioni}, {Vergani}, {Wang}, {Wang}, {Wang}, {Wang},
  {Watson}, {Webb}, {Wei}, {Willingale}, {Wu}, {Wu}, {Xin}, {Xu}, {Yu}, {Yu},
  {Yu}, {Zhang}, {Zhang}, {Zhang}, \& {Zhou}}]{wei2016}
{Wei}, J., {Cordier}, B., {Antier}, S., {et~al.} 2016, arXiv e-prints,
  arXiv:1610.06892, \dodoi{10.48550/arXiv.1610.06892}

\bibitem[{{Wei} \& {Wu}(2017)}]{2017IJMPD..2630002W}
{Wei}, J.-J., \& {Wu}, X.-F. 2017, International Journal of Modern Physics D,
  26, 1730002, \dodoi{10.1142/S0218271817300026}

\bibitem[{{Wei} {et~al.}(2014){Wei}, {Wu}, {Melia}, {Wei}, \&
  {Feng}}]{2014MNRAS.439.3329W}
{Wei}, J.-J., {Wu}, X.-F., {Melia}, F., {Wei}, D.-M., \& {Feng}, L.-L. 2014,
  \mnras, 439, 3329, \dodoi{10.1093/mnras/stu166}

\bibitem[{{White}(2020)}]{2020grbg.conf...51W}
{White}, N.~E. 2020, in Gamma-ray Bursts in the Gravitational Wave Era 2019,
  ed. T.~{Sakamoto}, M.~{Serino}, \& S.~{Sugita}, 51--53,
  \dodoi{10.48550/arXiv.2003.01592}

\bibitem[{{Woosley}(1993)}]{1993AAS...182.5505W}
{Woosley}, S.~E. 1993, in Bulletin of the American Astronomical Society,
  Vol.~25, American Astronomical Society Meeting Abstracts \#182, 894

\bibitem[{{Woosley} \& {Bloom}(2006)}]{2006ARA&A..44..507W}
{Woosley}, S.~E., \& {Bloom}, J.~S. 2006, \araa, 44, 507,
  \dodoi{10.1146/annurev.astro.43.072103.150558}

\bibitem[{{Yang} {et~al.}(2022){Yang}, {Ai}, {Zhang}, {Zhang}, {Liu}, {Wang},
  {Yang}, {Yin}, {Li}, \& {L{\"u}}}]{2022Natur.612..232Y}
{Yang}, J., {Ai}, S., {Zhang}, B.-B., {et~al.} 2022, \nat, 612, 232,
  \dodoi{10.1038/s41586-022-05403-8}

\bibitem[{{Yonetoku} {et~al.}(2004){Yonetoku}, {Murakami}, {Nakamura},
  {Yamazaki}, {Inoue}, \& {Ioka}}]{2004ApJ...609..935Y}
{Yonetoku}, D., {Murakami}, T., {Nakamura}, T., {et~al.} 2004, \apj, 609, 935,
  \dodoi{10.1086/421285}

\bibitem[{{Yonetoku} {et~al.}(2024){Yonetoku}, {Doi}, {Mihara}, {Matsuhara},
  {Sakamoto}, {Tsumura}, {Ioka}, {Arimoto}, {Enoto}, {Fujimoto}, {Goto},
  {Gunji}, {Hiraga}, {Ikunaga}, {Kawai}, {Kondo}, {Kurosawa}, {Li}, {Maeda},
  {Mitsuishi}, {Murakami}, {Nagataka}, {Nakagawa}, {Ogino}, {Owari}, {Sato},
  {Sato}, {Sato}, {Sawano}, {Serino}, {Shen}, {Sugita}, {Takahashi},
  {Tamagawa}, {Tamura}, {Tanaka}, {Tanimori}, {Tashiro}, {Togashi}, {Tomida},
  {Watanabe}, {Yamaoka}, {Yamauchi}, {Yatsu}, {Yoshida}, {Akitaya}, {Fukui},
  {Fukui}, {Ita}, {Kawabata}, {Matsuura}, {Miyasaka}, {Motohara}, {Narita},
  {Noda}, {Okita}, {Sano}, {Shinozaki}, {Tajima}, {Urata}, {Wada},
  {Yanagisawa}, {Yoshida}, {Bando}, {Jikuya}, {Minesugi}, {Miyazaki}, {Kono},
  {Takase}, {Nakatsubo}, {Kaga}, {Asano}, {Inayoshi}, {Inoue}, {Ito},
  {Izumiura}, {Kawanaka}, {Kinugawa}, {Kisaka}, {Kiuchi}, {Kyutoku},
  {Matsumoto}, {Mizuta}, {Murase}, {Nagakura}, {Nagataki}, {Nakada},
  {Nakamura}, {Niino}, {Suwa}, {Takahashi}, {Tanaka}, {Toma}, {Totani},
  {Yamazaki}, {Yokoyama}, {Harikane}, {Tanaka}, {Kimura}, \&
  {Kimura}}]{2024SPIE13093E..20Y}
{Yonetoku}, D., {Doi}, A., {Mihara}, T., {et~al.} 2024, in Society of
  Photo-Optical Instrumentation Engineers (SPIE) Conference Series, Vol. 13093,
  Space Telescopes and Instrumentation 2024: Ultraviolet to Gamma Ray, ed.
  J.-W.~A. {den Herder}, S.~{Nikzad}, \& K.~{Nakazawa}, 1309320,
  \dodoi{10.1117/12.3018571}

\bibitem[{{Yu} {et~al.}(2015){Yu}, {Wang}, {Dai}, \&
  {Cheng}}]{2015ApJS..218...13Y}
{Yu}, H., {Wang}, F.~Y., {Dai}, Z.~G., \& {Cheng}, K.~S. 2015, \apjs, 218, 13,
  \dodoi{10.1088/0067-0049/218/1/13}

\bibitem[{{Yuan} {et~al.}(2025){Yuan}, {Dai}, {Feng}, {Jin}, {Jonker},
  {Kuulkers}, {Liu}, {Nandra}, {O'Brien}, {Piro}, {Rau}, {Rea}, {Sanders},
  {Tao}, {Wang}, {Wu}, {Zhang}, {Zhang}, {Ai}, {Buchner}, {Bulbul}, {Chen},
  {Chen}, {Chen}, {Chen}, {Coleiro}, {Zelati}, {Dai}, {Fan}, {Fan},
  {Friedrich}, {Gao}, {Ge}, {Ge}, {Geng}, {Ghirlanda}, {Gianfagna}, {Gou},
  {Guillot}, {Hou}, {Hu}, {Huang}, {Ji}, {Jia}, {Komossa}, {Kong}, {Lan}, {Li},
  {Li}, {Li}, {Li}, {Li}, {Li}, {Ling}, {Liu}, {Liu}, {Liu}, {Liu}, {Luo},
  {Ma}, {Maggi}, {Maitra}, {Marino}, {Ng}, {Pan}, {Rukdee}, {Soria}, {Sun},
  {Tam}, {Thakur}, {Tian}, {Troja}, {Wang}, {Wang}, {Wang}, {Wei}, {Wen}, {Wu},
  {Wu}, {Xiao}, {Xu}, {Xu}, {Xu}, {Xu}, {Yang}, {You}, {Yu}, {Yu}, {Zhang},
  {Zhang}, {Zhang}, {Zhang}, {Zhang}, {Zhang}, {Zhou}, \&
  {Zou}}]{2025arXiv250107362Y}
{Yuan}, W., {Dai}, L., {Feng}, H., {et~al.} 2025, Science China Physics,
  Mechanics, and Astronomy, 68, 239501, \dodoi{10.1007/s11433-024-2600-3}

\bibitem[{{Y{\"u}ksel} {et~al.}(2008){Y{\"u}ksel}, {Kistler}, {Beacom}, \&
  {Hopkins}}]{2008ApJ...683L...5Y}
{Y{\"u}ksel}, H., {Kistler}, M.~D., {Beacom}, J.~F., \& {Hopkins}, A.~M. 2008,
  \apjl, 683, L5, \dodoi{10.1086/591449}

\end{thebibliography}

\end{document}